%% file: main.tex
\documentclass[10pt,journal,compsoc]{IEEEtran}

\usepackage{caption}
\usepackage{xspace} 
\usepackage{xcolor}
\usepackage{subcaption}
\usepackage{listings}
\usepackage{mdframed}
\usepackage{xspace}
\usepackage{footnote}
\usepackage{enumitem}
\usepackage{url}
\usepackage{booktabs} 
\usepackage[normalem]{ulem}
\usepackage{tcolorbox}
\usepackage{orcidlink}
\usepackage[numbers, sort]{natbib} 
\useunder{\uline}{\ul}{}
\newcounter{finding_counter}
\def\rqtwo{How do the \deep, \linevul, \reveal, and \ivdetect models perform in a realistic evaluation setting compared to the evaluation setting used in the original studies?\xspace}

\def\rqthree{How do the \deep, \linevul, \reveal, and \ivdetect models perform in a realistic evaluation setting when trained using a similar realistic training dataset?\xspace}

\def\BibTeX{{\rm B\kern-.05em{\sc i\kern-.025em b}\kern-.08em
    T\kern-.1667em\lower.7ex\hbox{E}\kern-.125emX}}
    
\AtBeginDocument{%
  \providecommand\BibTeX{{%
    \normalfont B\kern-0.5em{\scshape i\kern-0.25em b}\kern-0.8em\TeX}}}

\begin{document}

\newcommand{\reals}{{realistic setting}\xspace}

\newcommand{\linevul}{{LineVul}\xspace}

\newcommand{\reveal}{{ReVeal}\xspace}
\newcommand{\ivdetect}{{IVDetect}\xspace}
\newcommand{\nonvulnerable}{{uncertain}\xspace}
\newcommand{\deep}{{DeepWukong}\xspace}
\newcommand{\tbl}[1] { \textcolor{blue}{#1}}
\newcommand{\tre}[1] { \textcolor{red}{\textbf{#1}}}
\newcommand{\bigvul}{{Big-Vul}\xspace}

\newcommand{\dataset}{{\textit{Real-Vul}}\xspace}
\newcommand{\partha}[1] { \textcolor{red}{\textbf{[Partha: #1]}}}
\definecolor{my-gray}{cmyk}{0.01, 0, 0.01, 0.08, 1.00}
\newcommand{\boxtext}[1]{
    \begin{tcolorbox}[flushleft upper,boxrule=1pt,arc=1pt,left=0pt,right=0pt,top=0pt,bottom=0pt,colback=my-gray,after=\ignorespacesafterend\par\noindent]
        \textbf{ANSWER:} #1 \stepcounter{finding_counter}
    \end{tcolorbox}
}







\title{Revisiting the Performance of Deep Learning-Based Vulnerability Detection on Realistic Datasets}

\IEEEtitleabstractindextext{
\begin{abstract}
   \input{abstract}

\end{abstract}
}
\author{
    \IEEEauthorblockN{Partha Chakraborty\orcidlink{0000-0001-5965-615X}, \IEEEmembership{Student Member, IEEE}, Krishna Kanth Arumugam, \\Mahmoud Alfadel\orcidlink{0000-0002-2621-6104}, \IEEEmembership{Member, IEEE}, \\Meiyappan Nagappan\orcidlink{0000-0003-4533-4728}, and Shane McIntosh\orcidlink{0000-0002-0193-3975}, \IEEEmembership{Senior Member, IEEE}~\thanks{All authors are from David R. Cheriton School
of Computer Science, University of Waterloo, Canada. \\E-mail:\{p9chakra, kkarumug, malfadel, mei.nagappan, shane.mcintosh\}@uwaterloo.ca}}
}
\IEEEpubid{\copyright~2024 IEEE. Author pre-print copy. The final publication is available online at: \url{https://doi.org/10.1109/TSE.2024.3423712}}
\maketitle
\IEEEdisplaynontitleabstractindextext

\input{introduction}
\input{background}
\input{real-vul}
\input{experimental_design}
\input{preliminary_analysis}

\input{results}

\input{discussion}

\input{related_works}

\input{threats}
\input{implication}
\input{conclusion}

\def\UrlBreaks{\do\/\do-}
\bibliographystyle{IEEEtran}
\bibliography{bib.bib}
\vskip -2\baselineskip plus -1 fil
\begin{IEEEbiography}[{\includegraphics[width=1in,height=0.7in,clip,keepaspectratio]{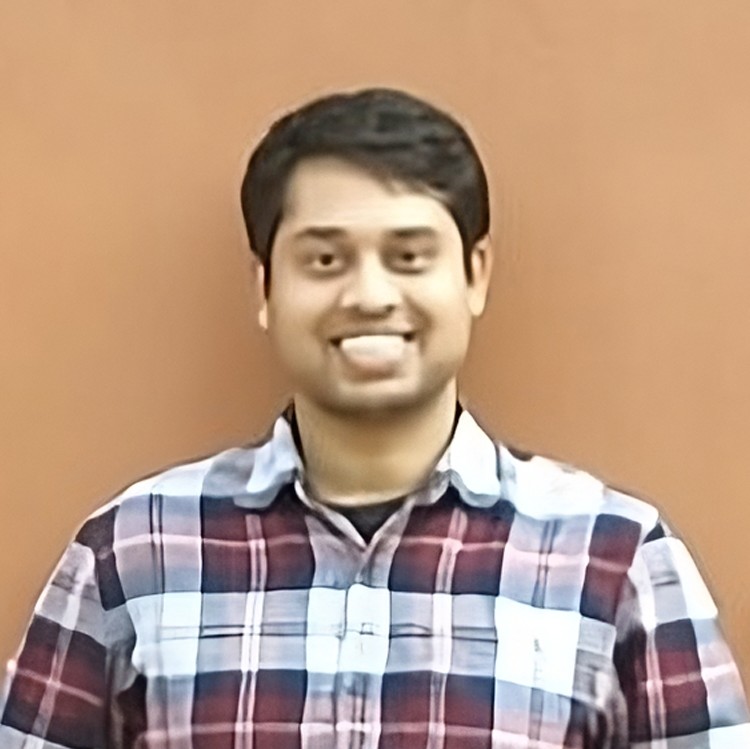}}]{Partha Chakraborty} is a Ph.D. candidate in the
Cheriton School of Computer Science at the
University of Waterloo, Canada. His research
interests include bug localization, vulnerability detection, and the use of machine learning techniques in software engineering. Find more about him at \url{https://parthac.me/.}
\end{IEEEbiography}
\vskip -4\baselineskip plus -1 fil
\begin{IEEEbiography}[{\includegraphics[width=1in,height=0.7in,clip,keepaspectratio]{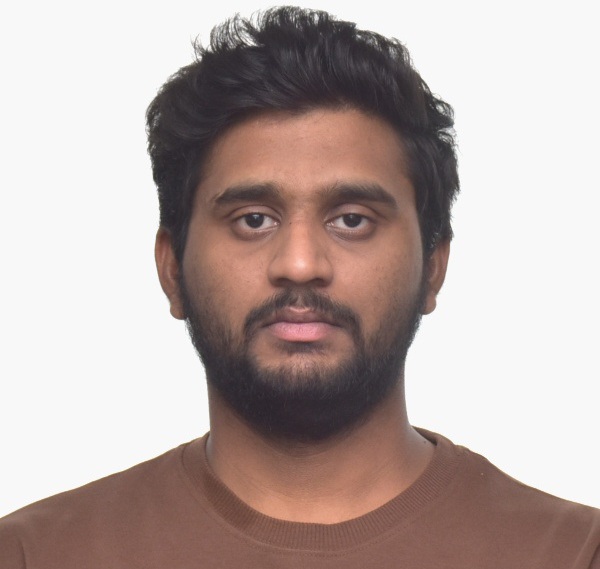}}]{Krishna Kanth Arumugam} was a Master’s student in the Cheriton School of Computer Science, University of Waterloo. His research interests include machine learning, security vulnerability detection, and mining software repositories.
\end{IEEEbiography}
\vskip -6\baselineskip plus -1 fil
\begin{IEEEbiography}[{\includegraphics[width=1in,height=0.7in,clip,keepaspectratio]{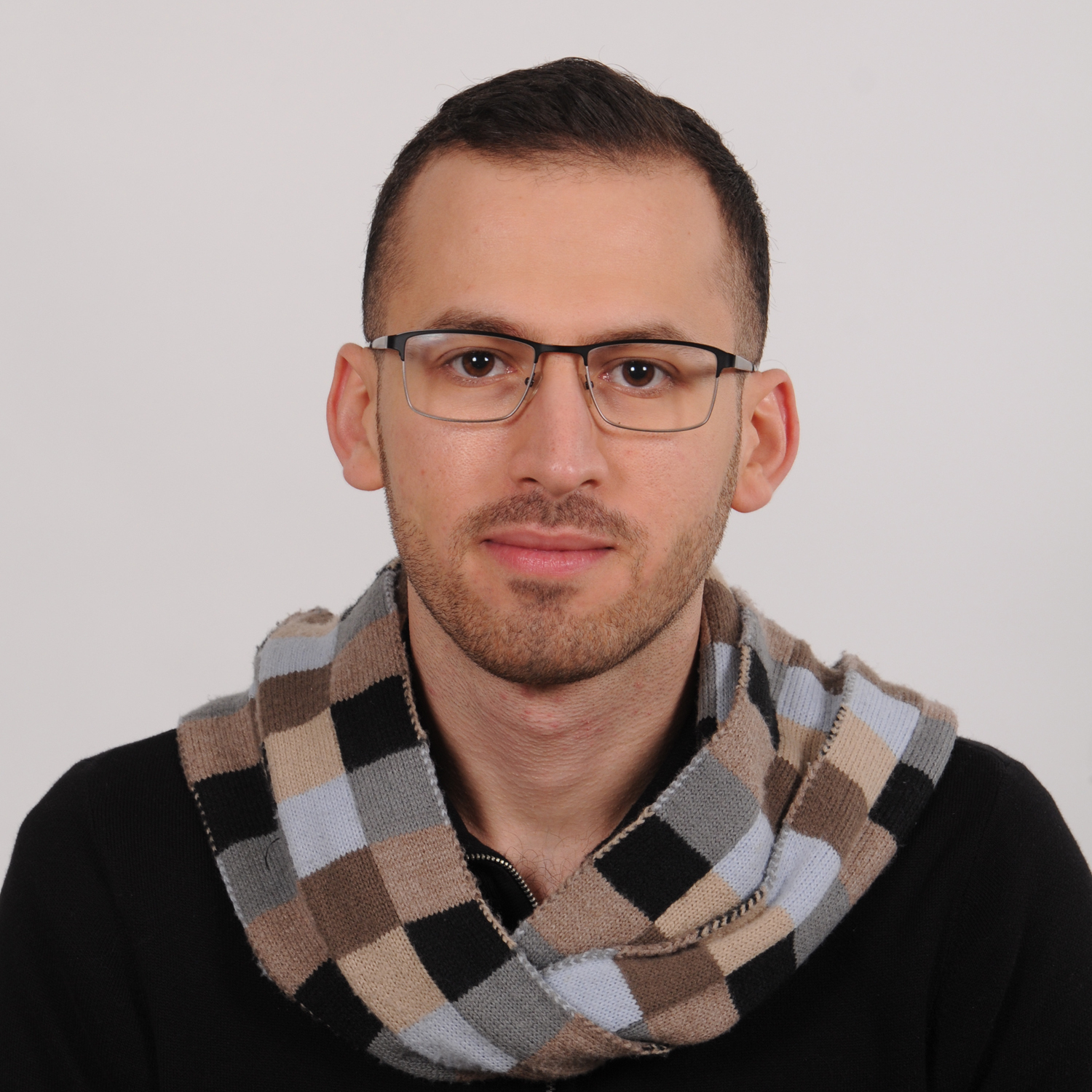}}]{Mahmoud Alfadel}  is a postdoctoral researcher
in the Cheriton School of Computer Science, University of Waterloo. His research interests include mining software repositories, empirical software engineering, software ecosystems, and release engineering. You can find
more about him at \url{https://rebels.cs.uwaterloo.ca/
member/mahmoud.html.}
\end{IEEEbiography}
\vskip -4\baselineskip plus -1fil
\begin{IEEEbiography}[{\includegraphics[width=1in,height=0.7in,clip,keepaspectratio]{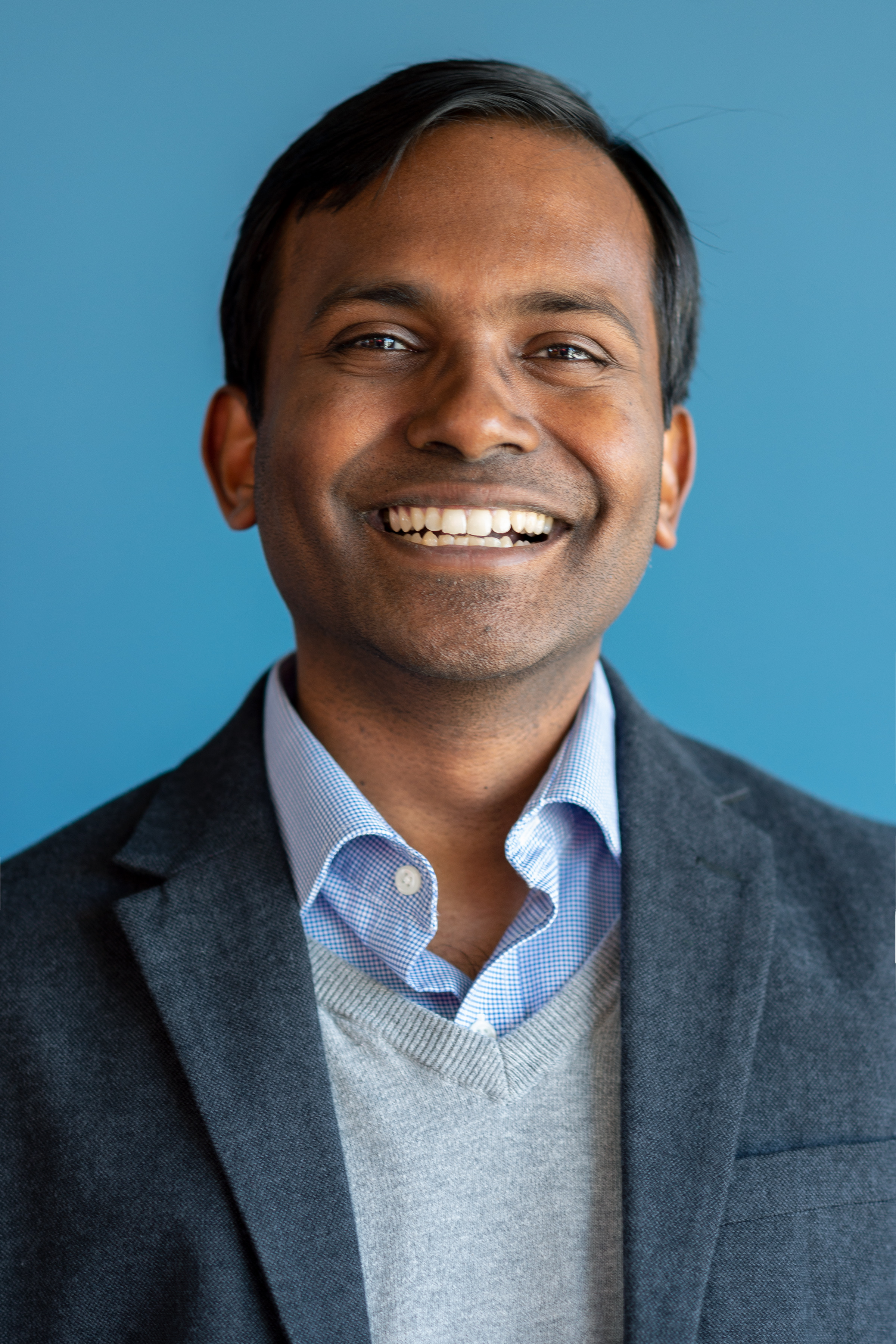}}]{Meiyappan Nagappan} is an Associate Professor at the Cheriton School of Computer Science, University of Waterloo. He has worked on empirical software engineering to address software development concerns and currently researches the impact of large language models on software development.
\end{IEEEbiography}
\vskip -5\baselineskip plus -1fil
\begin{IEEEbiography}[{\includegraphics[width=1in,height=0.7in,clip,keepaspectratio]{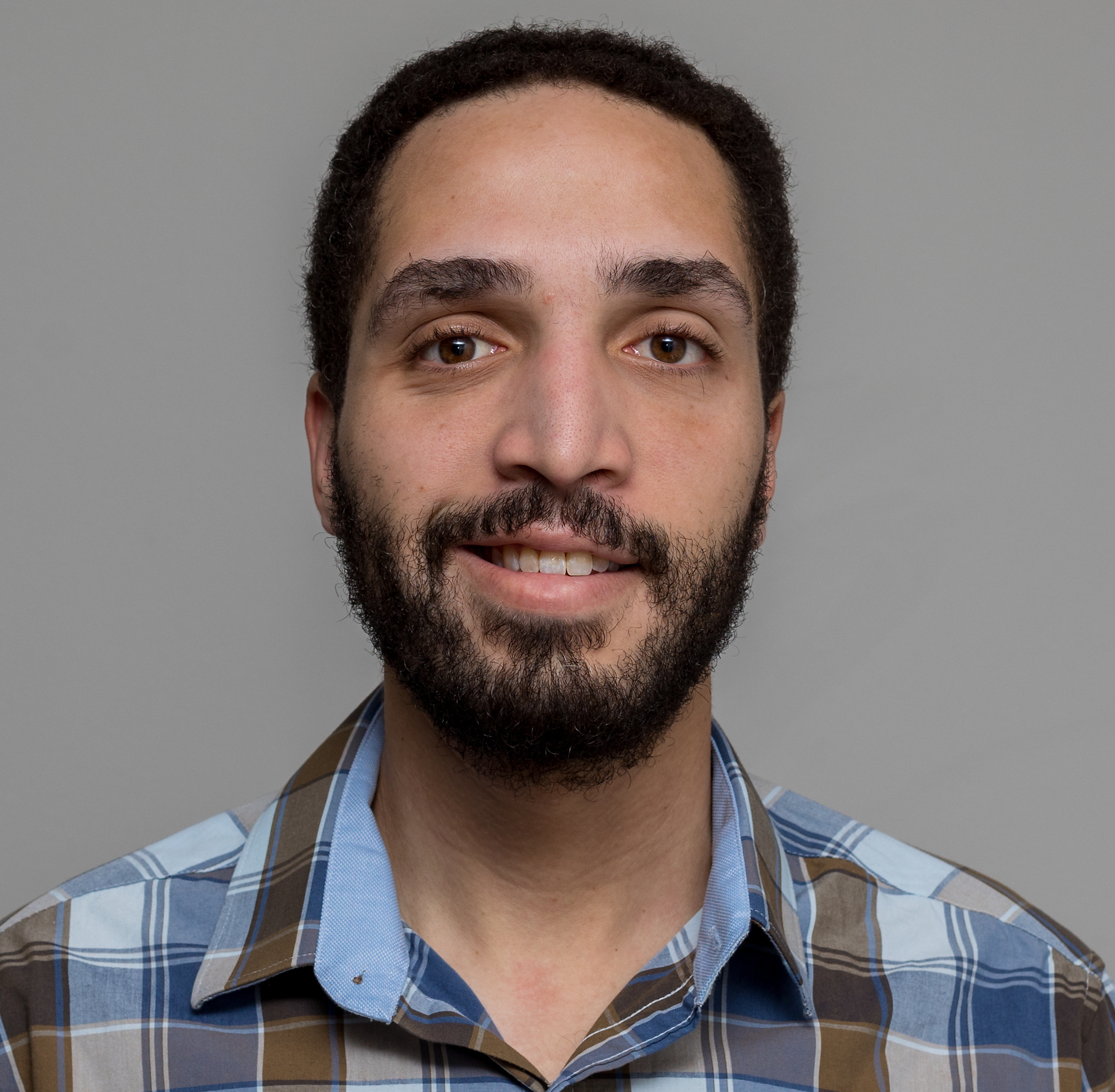}}]{Shane McIntosh} is an Associate Professor in the Cheriton School of Computer Science,  University of Waterloo, where he leads the Software Repository Excavation and Build Engineering Labs (Software REBELs). In his research, Shane uses empirical methods to study software build systems, devops pipelines, and software quality. More about Shane and his work can be found online at \url{https://rebels.cs.uwaterloo.ca/}
\end{IEEEbiography}
\end{document}

%% file: abstract.tex
The impact of software vulnerabilities on everyday software systems is concerning. Although deep learning-based models have been proposed for vulnerability detection, their reliability remains a significant concern. While prior evaluation of such models reports impressive recall/F1 scores of up to 99\%, we find that these models underperform in practical scenarios, particularly when evaluated on the entire codebases rather than only the fixing commit.
In this paper, we introduce a comprehensive dataset (\dataset) designed to accurately represent real-world scenarios for evaluating vulnerability detection models. We evaluate \deep, \linevul, \reveal, and \ivdetect vulnerability detection approaches and observe a surprisingly significant drop in performance, with precision declining by up to 95 percentage points and F1 scores dropping by up to 91 percentage points. 
A closer inspection reveals a substantial overlap in the embeddings generated by the models for vulnerable and \nonvulnerable samples (non-vulnerable or vulnerability not reported yet), which likely explains why we observe such a large increase in the quantity and rate of false positives. Additionally, we observe fluctuations in model performance based on vulnerability characteristics (e.g., vulnerability types and severity).
For example, the studied models achieve 26 percentage points better F1 scores when vulnerabilities are related to information leaks or code injection rather than when vulnerabilities are related to path resolution or predictable
return values.
Our results highlight the substantial performance gap that still needs to be bridged before deep learning-based vulnerability detection is ready for deployment in practical settings. We dive deeper into why models underperform in realistic settings and our investigation revealed overfitting as a key issue. We address this by introducing an augmentation technique, potentially improving performance by up to 30\%.
We contribute (a) an approach to creating a dataset that future research can use to improve the practicality of model evaluation; (b) \dataset -- a comprehensive dataset that adheres to this approach; and (c) empirical evidence that the deep learning-based models struggle to perform in a real-world setting.

%% file: introduction.tex
\section{Introduction}
Software vulnerabilities have a large negative impact on the software systems.
Identifying and addressing vulnerabilities in complex systems with multiple interconnected components has only exacerbated the problem, making a comprehensive and systematic approach necessary.
Machine learning models that are trained using Deep Neural Networks (DNNs) have shown promise in identifying software vulnerabilities~\cite{chakraborty2021deep,chakraborty2023rlocator}. 

However,  the reliability of these models in detecting vulnerabilities in real-world scenarios depends on the evaluation methodology and dataset. 
Biases that can affect model performance can arise from various sources, such as the manner in which the dataset is generated and labeled.
The generalizability of a model may suffer if the dataset on which it is trained is biased.
For instance, synthetic datasets, such as SARD~\cite{SARD_link}, are artificially created using fuzz techniques and evolutionary algorithms and are extensively used to evaluate the effectiveness of deep learning models in detecting software vulnerabilities \cite{cheng2021deepwukong, zheng2020impact,lin2020software}.
It is unclear the extent to which the vulnerabilities in these synthetic datasets capture the full range of complexities and variations that are found in real-world vulnerabilities. 

Moreover, these studies classified explicitly identified vulnerable functions as ``vulnerable'' and all remaining functions as ``non-vulnerable". However, this binary classification overlooks the possibility that some functions might be vulnerable yet undetected. Therefore, for the purposes of our study, we will describe these functions with a more nuanced term, ``\nonvulnerable," to acknowledge the inherent ambiguity in their vulnerability status.

On the other hand, there do exist real-world datasets that contain vulnerabilities from real software systems (e.g., \bigvul \cite{fan2020ac}, \reveal \cite{chakraborty2021deep}); however, these datasets tend to \textbf{(i)} only include a sample of code from a substantially larger codebase, i.e., the vulnerable and \nonvulnerable (non-vulnerable or vulnerability not reported yet) samples are extracted from vulnerability-related commits, and hence, the dataset does not reflect a realistic setting where vulnerability detection models are deployed to scan an entire project; and \textbf{(ii)} suffer from label inconsistency wherein the same sample is marked as both vulnerable and \nonvulnerable (see Section~\ref{datasets} for detailed explanation).
Consequently, we conjecture that models that are trained and evaluated using such datasets may not perform well when applied in real-world settings.


Therefore, in this paper, we introduce \dataset, a new vulnerability detection dataset designed to accurately represent the realistic settings in which these models would be deployed. By ``\textbf{realistic settings}," we refer to evaluation environments that closely mirror real-world conditions, including the distribution of vulnerable and uncertain samples. This approach ensures that the models are tested under conditions that reflect their actual usage scenarios.

\dataset differs from prior existing datasets in that it includes entire codebases in which vulnerable code exists, representing a more realistic setting in which vulnerability detection models would be applied. \dataset also accounts for the label inconsistency problem by comparing the hash of vulnerable and \nonvulnerable code segments.
%

To re-evaluate the performance of deep learning-based models in our more realistic setting,  we apply four \textit{state-of-the-art} models (i.e., \deep~\cite{cheng2021deepwukong}, \linevul~\cite{fu2022linevul}, \reveal~\cite{chakraborty2021deep}, and \ivdetect~\cite{Li2021}) to \dataset dataset. 
First, we train and evaluate these models on their original datasets, i.e., the SARD~\cite{SARD_link}, \bigvul~\cite{fan2020ac}, \reveal\cite{fan2020ac}, and \ivdetect~\cite{Li2021} datasets, respectively.
The \deep and \linevul models achieve notably high precision (87\% - 96\%) and F1  scores (90\% - 93\%), while \reveal and \ivdetect achieves precision and F1 scores of 29\% - 39\% and 78\% - 83\%, respectively.
However, when we evaluate these models on the \dataset dataset, we observe a substantial decrease in performance, e.g., precision and F1 score decreases of up to 95 and 91 percentage points, respectively.

To understand why these models produce a large number of false positives, we visualize their embeddings for vulnerable and \nonvulnerable samples in a two-dimensional space using t-SNE~\cite{van2008visualizing}. 
We find that these models struggle to establish a clear distinction between the two classes, resulting in inaccurate identification of vulnerabilities. For instance, models like \linevul heavily rely on lexical relationships between tokens and as such, encounter difficulties in accurately capturing the complex nature of vulnerabilities in real-world software.

To pinpoint why models underperform in realistic settings, we conducted an in-depth manual analysis of false positives. We found that overfitting, where models predict based on specific tokens, was a significant issue. To counteract this, we implemented an augmentation technique that has the potential to enhance performance by up to 30\%.
Furthermore, to understand the capability of the studied models in identifying different types of vulnerabilities, we stratify our model evaluation by vulnerability type.
Our findings indicate that the models achieve ~26 percentage points higher F1 scores in detecting vulnerability from specific types, such as information leaks and code injection, and struggle to detect vulnerabilities from other types, like path resolution and predictable return values.
Additionally, we find that the models perform poorly in identifying high-severity vulnerabilities, showing further challenges for these models in identifying important vulnerabilities.\\

\noindent
\textbf{Contributions.} This paper makes the following contributions:
\begin{itemize}[leftmargin=1em]
  \item We introduce \dataset, a new vulnerability detection dataset that aims to address the practical limitations of prior datasets.
  \item We re-evaluate the performance of existing \textit{state-of-the-art} vulnerability detection approaches on the \dataset dataset.
  Our evaluation provides a more accurate 
  representation of the performance of these models when used in realistic settings.
  \item We characterize the limitations of the studied approaches in terms of:  
  \begin{itemize}[leftmargin=1em]
      \item their embeddings, which show distinct overlaps across vulnerable and \nonvulnerable samples, suggesting models' inability to clearly distinguish vulnerable codes from \nonvulnerable codes.
    \item tendencies to be unable to detect vulnerabilities belonging to particular types and severities. Stratifying model performance by vulnerability type and severity allows targeted improvements and efficient resource allocation in addressing distinct security risks.
  \end{itemize}
\end{itemize}

\noindent
\textbf{Fostering open science:} To foster future advances to modeling approaches for vulnerability detection, we release our \dataset dataset, and we also make the scripts used for our experiments publicly available.\footnote{\url{https://zenodo.org/record/8206635 }\label{replication_package}}

%% file: background.tex
\section{Limitations of Existing Datasets}
\label{limitation}
As software systems become more complex and larger, the potential for security vulnerabilities also increases.
Therefore, it is crucial to have tools to discover these vulnerabilities.
Machine learning models have proven to be effective in understanding code and detecting vulnerabilities. However, obtaining and evaluating large datasets in a \reals~\cite{fan2020ac, zhou2019devign, chakraborty2021deep} remains challenging. 
Existing studies on vulnerability detection~\cite{fan2020ac, zhou2019devign} rely on datasets that are based on various criteria and may not reflect reality accurately. 
These datasets can be classified into three categories: synthetic, oracle-based, and real-world datasets.
In this section, we describe each category and their limitations.\\
\noindent
\textbf{Synthetic dataset (e.g., SARD ~\cite{SARD_link}).} 
 The Software Assurance Reference Dataset (SARD) is an example of a synthetic dataset. It contains a vast number of artificially produced C/C++ programs containing numerous security vulnerabilities. SARD has been created through automated methods like fuzzing or genetic algorithms. The SARD dataset's limitation lies in containing only artificially generated vulnerabilities, which may not accurately represent real-world software vulnerabilities. This dataset does not reflect the complexities and variations found in actual code written by developers, potentially limiting researchers from evaluating their models in a realistic setting.\\
\noindent
\textbf{Oracle-based dataset (e.g., D2A~\cite{zheng2021d2a}).} D2A is an example of an Oracle-based dataset. 
Unlike synthetic datasets, oracle-based datasets rely on third-party sources such as static analysis tools to provide labels for collected data samples. Although such datasets offer more complexity than synthetic datasets, they may not fully represent real-world vulnerabilities due to oversimplification. 
The inaccuracy of the labeling heavily affects the dataset's reliability which threatens the usability of the vulnerability detection models trained using this type of dataset.\\
\noindent
\textbf{Real-world dataset (e.g., 
\bigvul~\cite{fan2020ac}, Devign \cite{zhou2019devign}, \reveal~\cite{chakraborty2021deep}).} \bigvul  and \reveal  datasets are examples of  real-world datasets for vulnerability detection. They are more diverse than synthetic and oracle-based datasets; real-world datasets are generated using data available in issue-tracking systems and source code repositories.\par
\bigvul dataset \cite{fan2020ac} is composed of C/C++ functions collected from 348 open-source GitHub projects spanning from 2002 to 2019.
The Devign dataset has been curated from Linux Kernel, QEMU, Wireshark, and FFmpeg projects, whereas the \reveal dataset~\cite{chakraborty2021deep} is collected from the Chromium and Debian project. The above-mentioned datasets still have limitations.
First, such datasets fail to entirely capture vulnerability detection in a \reals where a comprehensive scan of the entire source code of a project would be performed. In \reals, the vulnerable functions are rare compared to \nonvulnerable ones, leading to an imbalance in their occurrence during scans. Conversely, prior datasets typically categorize altered or post-fix functions as \nonvulnerable and unchanged or pre-fix functions as vulnerable, resulting in an unrealistic nearly equal ratio of \nonvulnerable to vulnerable functions. Moreover, since vulnerable functions are rare in the real world, these datasets tend to be smaller and less diverse. Training on such a limited dataset will result in a biased model. Additionally, since the test dataset will also be biased, it will be challenging to detect these biases in the model.\par
\begin{figure}[tb!]
\centering
\includegraphics[width=8.5cm,keepaspectratio]{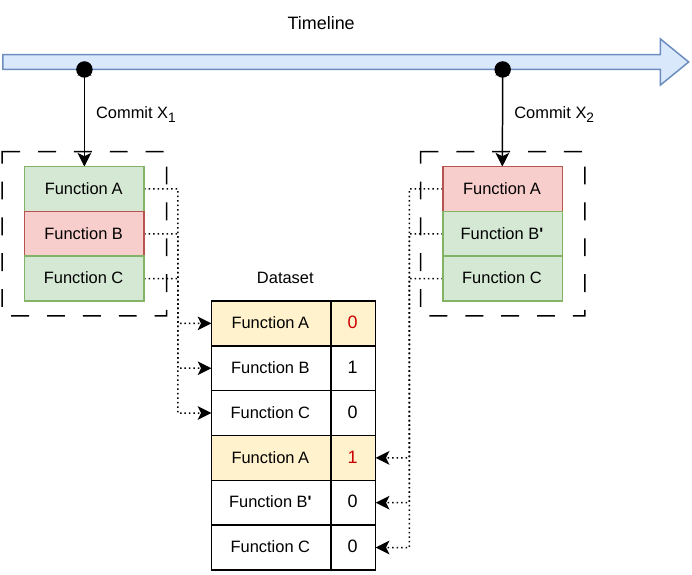}
\caption{Example of label inconsistency.}
\label{fig:label_inconsistency}
\end{figure}

Second, these datasets may suffer from \emph{label inconsistency}. Figure~\ref{fig:label_inconsistency} elaborates further on the problem of label inconsistency. 
In the figure, consider two vulnerability-fixing commits, $X_1$ and $X_2$, with $X_1$ occurring before $X_2$. 
In $X_1$, the vulnerable function is Function B (note that after fixing it, it became Function B\textquotesingle).
Following the data collection policy of previous studies (e.g., \reveal~\cite{chakraborty2021deep}), Function B would be labeled as vulnerable, while the other unchanged functions (Function A and C) would be labeled as \nonvulnerable. However, as we progress further in the timeline, we encounter commit $X_2$, which fixes the vulnerability in "Function A". Now, as per the aforementioned policy, "Function A" would be labeled as vulnerable, while Function B\textquotesingle and C would be labeled as \nonvulnerable.
Consequently, the dataset will contain two entries for Function A, where in one case, it is marked as vulnerable and in the other as \nonvulnerable. 
This discrepancy in labeling creates the label inconsistency issue.
In fact, we verified the presence of label inconsistency in existing datasets and found that 15\% vulnerable samples in the \reveal dataset have been listed as \nonvulnerable samples.\par
Due to the aforementioned limitations, current deep learning models for vulnerability detection (e.g., ~\cite{chakraborty2021deep, fu2022linevul, cheng2021deepwukong, li2018vuldeepecker, zhou2019devign}) could result in unrealistic evaluations compared to real-world usage.
Additionally, the reported model performance might be either exaggerated or understated.
Therefore, in this study, we build \dataset, a new dataset that takes into consideration (1) a realistic setting where a comprehensive scan of the entire source
code of a project would be performed; and (2) label consistency.
The next section presents \dataset and the process we follow to create it.
%

%% file: real-vul.tex
\section{\dataset Dataset}
\label{datasets}
In this section, we introduce our proposed dataset, named \dataset, which addresses the limitations of existing vulnerability detection datasets (Section~\ref{limitation}). Table~\ref{tab:dataset} provides an overview of the projects that are incorporated in \dataset.\\ 
\input{tables/dataset_information}
\noindent
\textbf{Project selection.}
\dataset includes separate training and testing datasets designed for training and evaluating vulnerability detection models. To ensure high-quality vulnerable samples, we carefully select projects from the 
\bigvul dataset~\cite{fan2020ac} based on the number of vulnerabilities in each project. \bigvul dataset comprises real-world vulnerable C/C++ functions. It also provides rich metadata for vulnerabilities, including line numbers, vulnerability-fixing commit hashes,  CVE IDs, severity rankings, and summaries from public Common Vulnerabilities and Exposures (CVE) databases and related source code repositories. Consequently, all samples within the \dataset dataset are real and written in C/C++ as well. Our focus on projects with a higher number of vulnerabilities signifies the substantial effort invested in identifying vulnerabilities in these projects~\cite{Walden2020}. Our selection consists of the top 10 projects with the most vulnerabilities, creating a dataset that is at least two times larger than those used in previous studies~\cite{chakraborty2021deep, zhou2019devign}. Remarkably, these ten projects encompass 73\% of the vulnerable samples from the entire \bigvul dataset, making \dataset a representative subset without its limitations.\\
\noindent
\textbf{Creation of vulnerable samples.} To align with real-world scenarios, we adopt a time-based strategy for sample creation in the training and testing datasets. This approach simulates the process of training models on historical data and then identifying vulnerabilities over time. The time-based strategy is described below:

For generating vulnerable samples, we start extracting the dates on which the vulnerable functions were fixed using the vulnerability-fixing commit hashes available in the \bigvul dataset.

To organize vulnerable samples, we order the functions based on their vulnerability-fixing dates (i.e., the dates of vulnerability-fixing commits). We then allocate the first 80\% of the ordered vulnerable functions to the training dataset, while the remaining 20\% form the test dataset. For example, in the case of the FFmpeg project, the training dataset includes samples fixed between August 3, 2013, and May 30, 2018. The testing set comprises samples fixed between June 28, 2018, and August 5, 2019. 
Note that these dates are the vulnerability-fixing commit dates in the FFmpeg project. The dates for all of the projects are available in the online Appendix.
In total, our dataset consists of 5,528 vulnerable functions, with 4,418 functions assigned to the training dataset and 1,110 to the test dataset.\\
\noindent
\textbf{Creation of \nonvulnerable samples.}
To create \nonvulnerable samples, we clone the remote repository of each selected project and check out two snapshots, one for the training dataset and one for the testing dataset, using the most recent snapshots for each. For instance, in the FFmpeg project example, we clone the FFmpeg repository on May 31, 2018, to generate \nonvulnerable samples for the training dataset. This ensures that the \nonvulnerable samples are from a version later than the last vulnerability-fixing date in the training dataset (May 30, 2018). For the test dataset, we take the second snapshot on August 6, 2019.
It is essential to maintain the constraint of chronological dates for training and testing snapshots.
That said, other snapshots of the project could be taken as long as this constraint is upheld.

To obtain \nonvulnerable functions, we extract all functions from the source code files collected during snapshot cloning. We calculate the hash of these collected functions and the previously gathered vulnerable functions. Functions whose MD5 hashes do not match any of the MD5 hashes of the vulnerable functions are labeled as \nonvulnerable, ensuring label consistency in the \dataset dataset.

Upon completion, we have a total of 1,682,713 \nonvulnerable functions, with 769,464 allocated to the training dataset and 913,249 to the test dataset. The higher number of \nonvulnerable functions in the test dataset is due to their extraction from a later project version compared to the one used for the training dataset.

%% file: tables/dataset_information.tex
\begin{table*}[tb]
\caption{Projects included in \dataset dataset.}
\label{tab:dataset}
\centering
\begin{tabular}{ccrrr}
\toprule
\multicolumn{1}{c}{\textbf{Project}} & \multicolumn{1}{c}{\textbf{Description}} & \multicolumn{1}{p{2cm}}{\textbf{Project Size (\# of Functions)}} & \multicolumn{1}{p{2cm}}{\textbf{\# Vulnerable Functions}} & \multicolumn{1}{p{2.5cm}}{\textbf{\# Uncertain Functions}} \\ \midrule
Chromium                              & Open source browser                       & 153,057                                                      & 3,137                                                 & 149,920                                                  \\ \midrule
FFmpeg                                & Video and audio processing                & 7,071                                                        & 85                                                    & 6,986                                                    \\ \midrule
ImageMagic                            & Image manipulation tool                   & 1,401                                                        & 201                                                   & 1,200                                                    \\ \midrule
Jasper                                & Image encoding tool                       & 315                                                          & 86                                                    & 229                                                      \\ \midrule
Krb5                                  & Computer network authentication protocol  & 3,151                                                        & 106                                                   & 3,045                                                    \\ \midrule
Linux                                 & Operating system                          & 91,392                                                       & 1,477                                                 & 89,915                                                   \\ \midrule
Openssl                               & Encryption tool suit                      & 2,568                                                        & 110                                                   & 2,458                                                    \\ \midrule
Php-src                               & Interpreter for php-scripting language    & 3,613                                                        & 104                                                   & 3,509                                                    \\ \midrule
Qemu                                  & System emulator                           & 7,739                                                        & 82                                                    & 7,657                                                    \\ \midrule
Tcpdump                               & Computer network analyzing tool           & 612                                                          & 140                                                   & 472                                                      \\ \bottomrule
\end{tabular}
\end{table*}

%% file: experimental_design.tex
\section{Study Design}
In this section, 
we present the models we evaluate using \dataset dataset (Section~\ref{models}).
Then, we describe the research questions driving our investigation (Section~\ref{research_questions}), and the evaluation metrics (Section~\ref{evaluation_metrics}).

\subsection{Employed Models}
\label{models}
We choose four \textit{state-of-the-art} deep learning-based models, namely \linevul~\cite{fu2022linevul}, \deep \cite{cheng2021deepwukong}, \reveal~\cite{chakraborty2021deep}, and \ivdetect~\cite{Li2021} to conduct our experiments. 
Next, we briefly describe the architectural details of these models.\\
\noindent
\textbf{\linevul.}  \linevul \cite{fu2022linevul} is a deep learning-based model built using CodeBERT~\cite{feng2020codebert}. 
We opt to include \linevul in our analysis as
it is the state-of-the-art deep learning-based vulnerability detection model that utilizes a sequence-based approach for vulnerability detection. \linevul takes in a chunk as input and classifies it as a vulnerable or \nonvulnerable \textit{chunk} utilizing the CodeBERT model. A chunk is a sequence of code tokens generated from source code programs. The \linevul has been trained on the \bigvul dataset which has been curated using vulnerability-fixing commits from open-source projects.\\
\noindent
\textbf{\deep.} \deep \cite{cheng2021deepwukong} is a leading vulnerability detection model utilizing a Graph Neural Network (GNN)~\cite{cangea2018towards} for in-depth code analysis. It converts code into Program Dependence Graphs (PDGs), then into XFGs (subgraphs), highlighting the data and control flow dependencies within. \deep evaluates these XFGs to identify vulnerable code segments, employing the SARD dataset for training. This model has been included in the study due to its advanced graph-based approach, which sets a new benchmark for accurately detecting code vulnerabilities.\\
\noindent
\textbf{\reveal.}  \reveal~\cite{chakraborty2021deep} is a model that finds vulnerable code using graph neural networks. It has been chosen because it gathers data similarly to \dataset, focusing on unchanged functions from specific commits as non-vulnerable examples. \reveal combines control flow, data flow, syntax trees, and dependency graphs into a Code Property Graph (CPG) for comprehensive code analysis. It trains on a dataset curated from the Linux Debian Kernel and Chromium project.\\
\noindent\textbf{\ivdetect.} \ivdetect~\cite{Li2021} is a tool that aims to provide precise interpretations of detected vulnerabilities. \ivdetect incorporates representation learning and a graph-based interpretation model. It processes code by analyzing control flow, data flow, abstract syntax trees, and program dependency graphs, creating a unified structure for efficient vulnerability detection. \ivdetect uses datasets created by prior studies~\cite{chakraborty2021deep, fan2020ac} comprising projects such as FFmpeg, Qemu, and Chromium.
\subsection{Evaluation Metrics.} \label{evaluation_metrics}
The evaluation metric must accurately measure the models' performance on the task at hand. In this study, we use accuracy, precision, recall, F1, and AUC to measure the models' performance.


Accuracy quantifies the overall correctness of a vulnerability detection model, representing the proportion of true results (both true positives and true negatives) among the total number of samples that were evaluated. For example, if a model correctly identifies 80 out of 100 functions (vulnerable or not), its accuracy is 80\%. High accuracy indicates effective identification of both vulnerable and \nonvulnerable samples with minimal errors.

Precision represents the fraction of vulnerabilities detected by the model that are truly vulnerabilities. 
For example, in the case of the \deep model, it measures how many XFGs predicted as vulnerable are genuinely vulnerable XFGs.
A low precision score indicates that the model is incorrectly classifying many \nonvulnerable samples as vulnerable, leading to a high number of false positives. 
Conversely, a high precision score implies that when the model detects a vulnerability, it is likely a real vulnerability.

Recall, on the other hand, denotes the fraction of actual vulnerabilities in the system that the model successfully detects. 
For the \deep model, it measures how many true vulnerable XFGs are correctly identified as vulnerable by the model. 
A high recall score indicates that the model can detect most vulnerabilities correctly, resulting in a low number of false negatives.

The F1 score combines precision and recall, providing an overall assessment of how well these two measurements are balanced. 
By considering both precision and recall, the F1-score offers a more complete understanding of the model's effectiveness in vulnerability detection.

Area Under the Curve (AUC) evaluates a model's capability to distinguish between classes, like differentiating vulnerable from \nonvulnerable XFGs in the \deep model. A high AUC score indicates that the model is proficient in accurately distinguishing between vulnerable and \nonvulnerable XFGs. Notably, in a balanced dataset, a random model would achieve an AUC of 0.50.

\subsection{Research Questions}
\label{research_questions}
We introduce two research questions (RQs) and explain the motivation behind each one.\\
\noindent
\textbf{RQ$_1$:} \textit{\rqtwo}\\
\deep, \linevul, \reveal, and \ivdetect are four state-of-the-art vulnerability detection models that have demonstrated promising results in identifying security vulnerabilities. However, we contend that the datasets used to evaluate these models do not accurately represent real-world usage, where the models would scan entire project source code files (see Section~\ref{limitation} for a comprehensive overview of dataset limitations).
For instance, the \deep model heavily relies on the SARD dataset for evaluation, which contains artificially generated samples lacking the complexities of real-world vulnerabilities. Similarly, the \linevul model used the \bigvul dataset \cite{fan2020ac}, which suffers from a scarcity of \nonvulnerable samples. Additionally, the \reveal dataset also suffers from limited \nonvulnerable samples and significant ``label-inconsistency.''
In this RQ, we evaluate the \deep, \linevul, \reveal, and \ivdetect models using our \dataset test dataset, which provides a more realistic representation of vulnerability detection model performance in practical scenarios.\\
\noindent
\textbf{RQ$_2$:} \textit{\rqthree}\\
In RQ$_1$, we utilize the \dataset dataset for model evaluation. However, we recognize that the training dataset employed to train these models might not accurately represent the same distribution as the evaluation dataset (i.e., \dataset). 
Doing so can lead to poor model performance as the training and testing datasets are from different distributions.
To achieve optimal results, it is essential for both datasets to share the same distribution as the data encountered in practical usage. Consequently, in RQ$_2$, we investigate the impact of data distribution on model performance by training and testing the models using our \dataset dataset. 
This exploration will enable us to ascertain whether employing a training dataset representative of the same distribution as the evaluation dataset leads to enhanced model performance.

%% file: preliminary_analysis.tex
\section{Preliminary Analysis}
\label{sec:replication}
Our study aims to evaluate the performance of the \deep, \linevul, and \reveal models in a more realistic setting.
As a first step towards our goal, we run the \deep, \linevul, \reveal, and \ivdetect models on the SARD, the \bigvul, the \reveal, and the \ivdetect datasets, respectively, and verify the findings reported in the original papers.
By doing so, we aim to ensure that the results previously reported are reliable and can be replicated.
Additionally, we use the results obtained from these experiments as our baseline models, which we compare to the performance of the models in our introduced realistic settings.\par
\noindent
\textbf{Approach.}
To evaluate the \deep, \linevul, \reveal, and \ivdetect models on the dataset used in their studies, we first download the SARD dataset,\footnote{https://github.com/jumormt/\deep} the \bigvul dataset,\footnote{https://github.com/awsm-research/\linevul} the \reveal dataset,\footnote{https://drive.google.com/drive/folders/1KuIYgFcvWUXheDhT--cBALsfy1I4utOy\reveal}  used in \deep \cite{cheng2021deepwukong}, \linevul \cite{fu2022linevul}, \reveal \cite{chakraborty2021deep}, and \ivdetect~\cite{Li2021} studies, respectively. 
We create the \deep model inputs (XFGs), the \linevul model inputs (chunks), the \reveal model inputs (CPG), and the \ivdetect models input PDG using the SARD dataset,  the \bigvul dataset, the \reveal dataset, and the \ivdetect dataset samples, respectively.
The generated XFG, chunk, CPG, and PDG datasets are split into training and testing datasets where 80\% of the samples belong to the training dataset and the remaining 20\% samples belong to the testing dataset. 
Following the authors' original experiments, we train the \deep model for 50 epochs and the \linevul model for ten epochs. 
For the \reveal and \ivdetect model, we follow the original implementation of \reveal and \ivdetect. 
The maximum epoch is set to 100 and 50 for \reveal and \ivdetect, respectively.  
We stop the training procedure if the F1-score does not increase for five consecutive epochs. 
Finally, We evaluate the models using the testing datasets.\\
\input{tables/replication_result}

\noindent\textbf{Results.}
Table~\ref{table:RQ0} shows the replication results of the \deep, \linevul, \reveal and \ivdetect models. 
From the figure, we observe that the \deep model achieves 98\%, 87\%, 98\%, 93\%, and 88\% for accuracy, precision, recall, F1-score, and AUC, respectively. 
The accuracy and F1-score for the \deep model differ by +1 and -2 percentage points, respectively, from the results reported by the \deep paper.

Our results for the \linevul model follow a similar trend.
The accuracy, precision, recall, F1-score, and AUC for the \linevul model are 96\%, 96\%, 84\%, 90\%, and 85\% respectively. 
Comparing with the results reported in the  \linevul paper, we find that the precision, recall, and  F1-score differ by -1, -2, and -1 percentage points, respectively.

The \reveal model demonstrates an accuracy of 81\%, with precision, recall, F1-score, and AUC values of 29\%, 59\%, 38\%, and 78\%, respectively. Upon comparing these results with the findings reported in the \reveal paper, we note a difference of -6, -3, and -7 percentage points in precision, recall, and F1-score, respectively.

The \ivdetect model demonstrates an accuracy of 85\%, with precision, recall, F1-score, and AUC values of 39\%, 63\%, 24\%, and 83\% respectively. Upon comparing these results with the findings reported in the \ivdetect paper, we note a difference of -2, -2, and -24 percentage points in precision, recall, and F1-score, respectively.

\begin{figure*}[tb!]
  \centering
  \begin{subfigure}[t]{0.24\textwidth}
    \centering
    {\includegraphics[width=\textwidth]{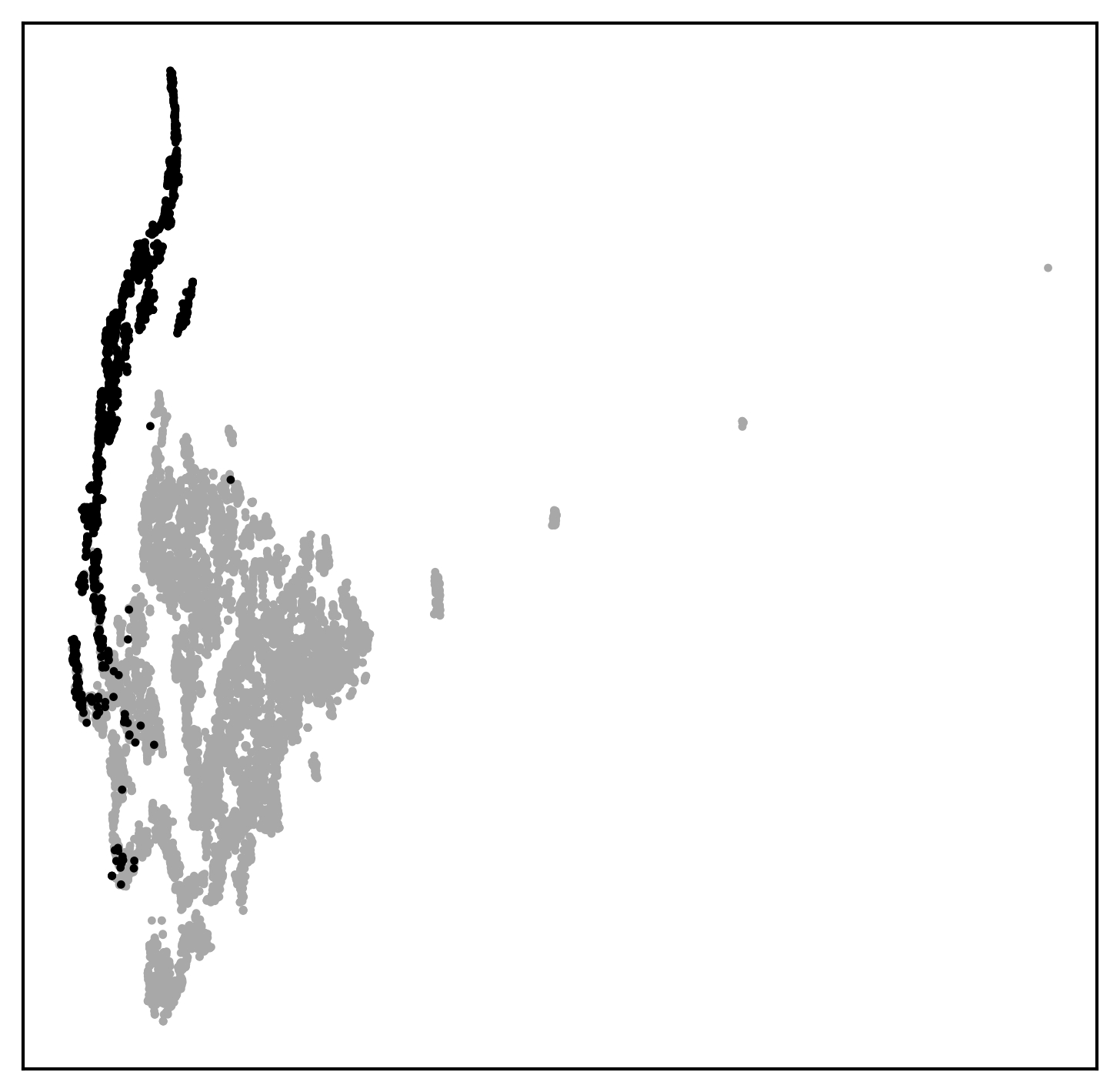}}
    \caption{\deep.}
    \label{fig:tsneplots0}
  \end{subfigure}
  \begin{subfigure}[t]{0.24\textwidth}
    \centering
    \includegraphics[width=\textwidth]{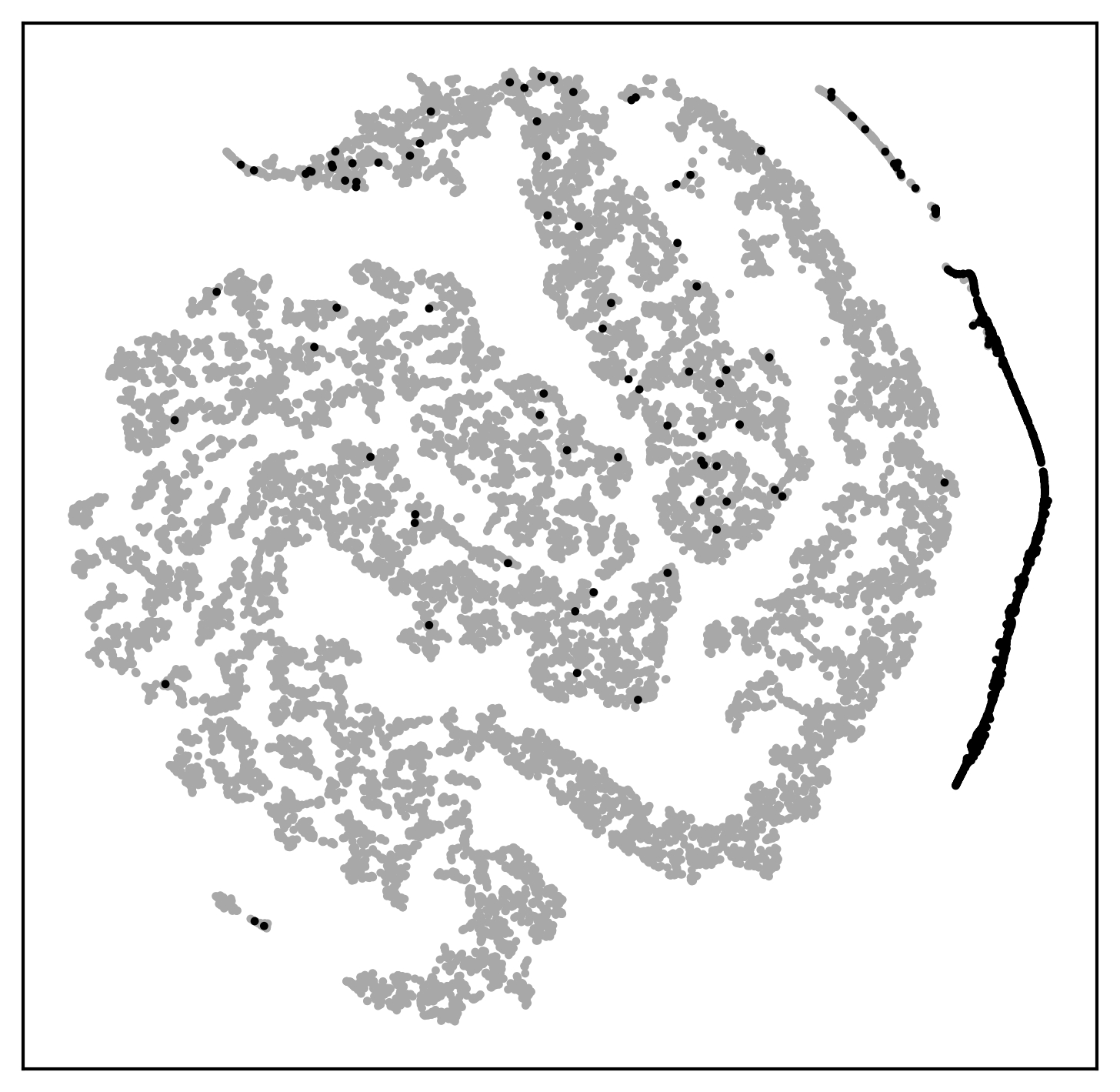}
    \caption{\linevul.}
    \label{fig:tsneplots1}
  \end{subfigure}
  \begin{subfigure}[t]{0.233\textwidth}
    \centering
    \includegraphics[width=\textwidth]{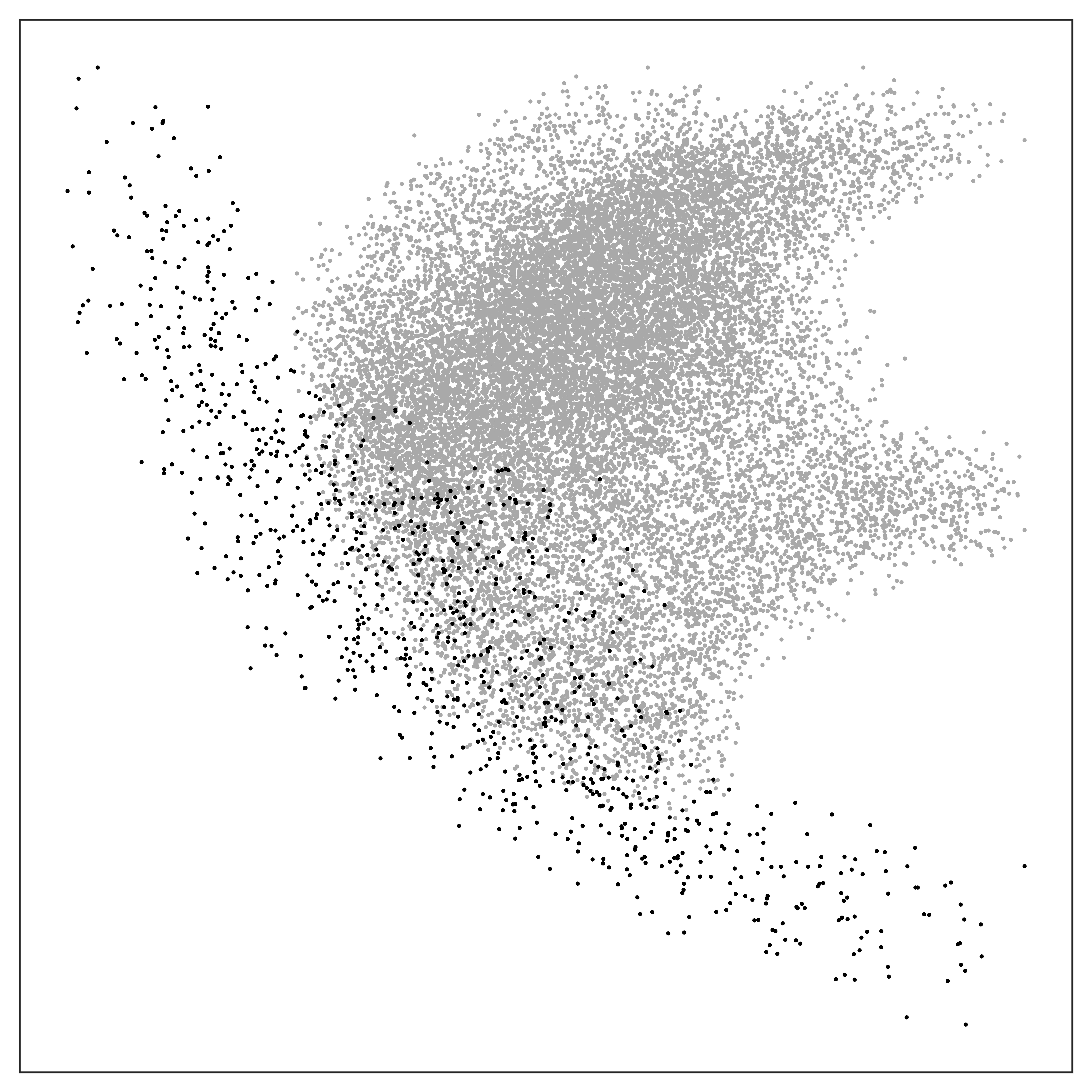}
    \caption{\reveal.}
    \label{fig:tsneplots2}
  \end{subfigure}
  \begin{subfigure}[t]{0.233\textwidth}
    \centering
    \includegraphics[width=\textwidth]{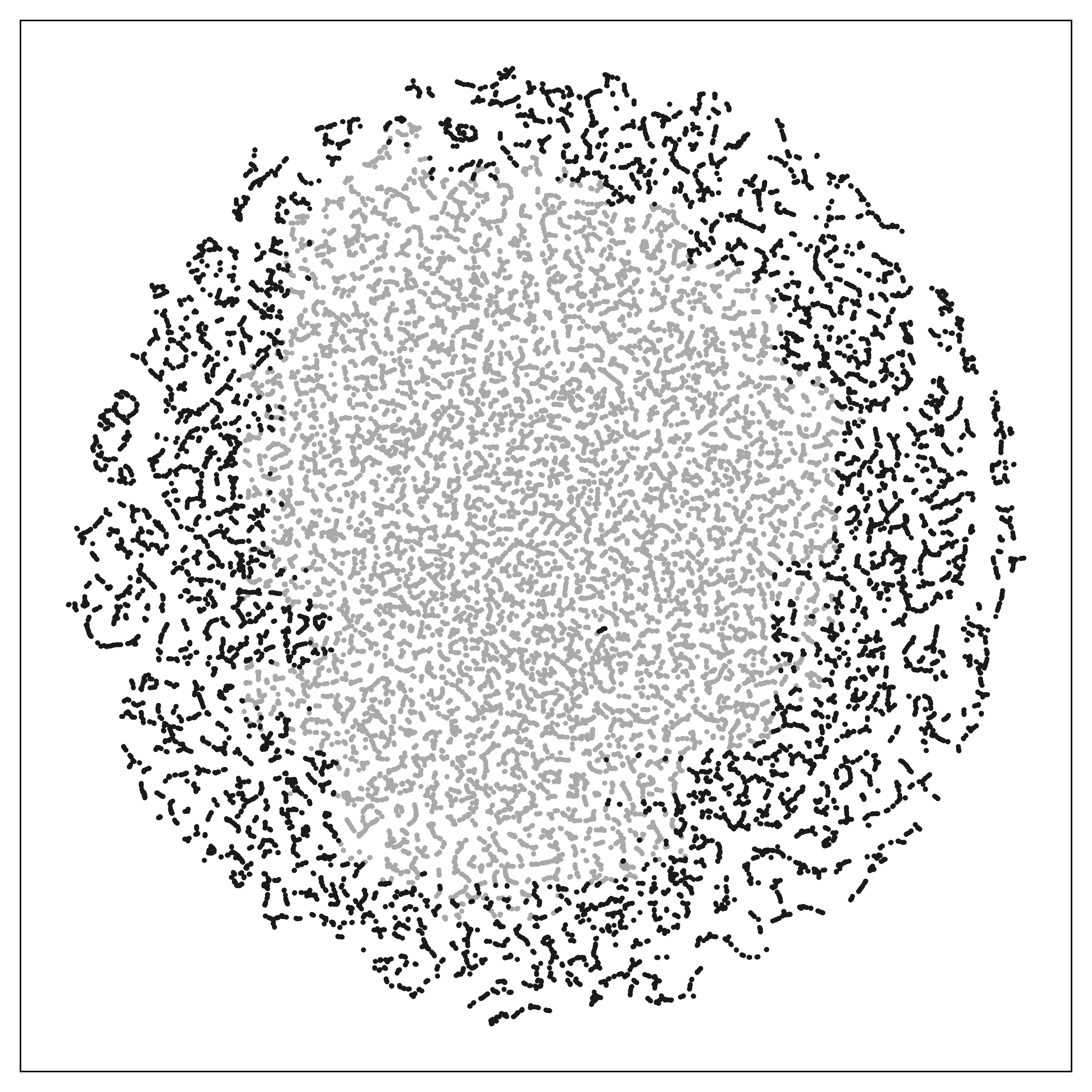}
    \caption{\ivdetect.}
    \label{fig:tsneplots3}
  \end{subfigure}
   \caption{Scatter plots showing the class separation between vulnerable and \nonvulnerable samples in the original datasets of the approaches. \raisebox{-0.4ex}{\textcolor{black}{\scalebox{2}{\textbullet}}} denotes vulnerable samples and \raisebox{-0.4ex}{\textcolor{gray}{\scalebox{2}{\textbullet}}} denotes \nonvulnerable samples.}
\label{tsneplotsRQ1}
\end{figure*}
Overall, we observe that the differences in the metrics are negligible. 
The little difference in the metrics can be because of the presence of different samples in the training/testing datasets. 
When a dataset is split into a training set and a testing set, the samples in the two sets are chosen randomly. 
This means that the samples in the training set and the testing set will be different each time we split the dataset. 
As a result, model performance on the testing set may vary from one split to another. 
The model performance on the testing set may depend on the specific samples that are included in the testing set. 
For example, if the testing set contains a particularly difficult or easy sample, this can affect the overall performance.
These results serve as baseline models for comparison with the performance of the models tested under realistic settings in the RQs.

The effectiveness of these models in vulnerability detection is influenced by the degree to which the embeddings of the vulnerable and \nonvulnerable classes are distinct and separable. 
That is, the greater the distinction and separability of the embedding, the more straightforward it is for the model to differentiate between the two classes. 
To explore this aspect, we conduct an experiment that involves visualizing the model's capability to distinctly segregate samples belonging to the vulnerable and \nonvulnerable classes.

We employ t-distributed Stochastic Neighbor Embedding (t-SNE) \cite{van2008visualizing} to visualize the embeddings produced by the \deep, \linevul, \reveal, and \ivdetect models. T-SNE is a machine learning algorithm for visualizing high-dimensional data in a low-dimensional (typically two or three) space. It calculates similarities between data points, converting them into probabilities and minimizing their Kullback-Leibler divergence~\cite{Joyce2011} to preserve local structures. This process effectively clusters similar data, which makes t-SNE valuable for identifying patterns and relationships in multi-dimensional data.

We extract the embedding from the \deep, \linevul, \reveal, and \ivdetect models used in this RQ. 
The \textit{[CLS]} embedding vectors generated by the CodeBERT model represent the embedding for the \linevul model, while the final fixed vector produced by the Graph Neural Network and the representation-learning model represents the embedding for the \deep, \reveal, \ivdetect model, respectively. It is worth noting that the embeddings are generated for the testing datasets.

We create four scatter plots in total, each displaying the reduced embedding along with their corresponding labels (vulnerable or \nonvulnerable). 
The scatter plots are presented in Figure~\ref{tsneplotsRQ1}.
We observe that all the models we created, except \reveal and \ivdetect, exhibit clear separation between vulnerable and \nonvulnerable samples. 
The inability to clear separation may explain the comparatively lower performance of the \reveal and \ivdetect model.

%% file: tables/replication_result.tex
\begin{table}[tb]
\caption{Replication results of \deep, \linevul, \reveal, and \ivdetect models.}
\label{table:RQ0}
\centering
\begin{tabular}{@{}crrrrr@{}}
\toprule
\textbf{Model} & \multicolumn{1}{c}{\textbf{\begin{tabular}[c]{@{}c@{}}Accuracy \\ (\%)\end{tabular}}} & \multicolumn{1}{c}{\textbf{\begin{tabular}[c]{@{}c@{}}Precision \\ (\%)\end{tabular}}} & \multicolumn{1}{c}{\textbf{\begin{tabular}[c]{@{}c@{}}Recall \\ (\%)\end{tabular}}} & \multicolumn{1}{c}{\textbf{\begin{tabular}[c]{@{}c@{}}F1\\ (\%)\end{tabular}}} & \multicolumn{1}{c}{\textbf{\begin{tabular}[c]{@{}c@{}}AUC\\ (\%)\end{tabular}}} \\ \midrule
\deep     & 98                                    & 87                                     & 98                                  & 93                              & 88                               \\ \midrule
\linevul        & 96                                    & 96                                     & 84                                  & 90                              & 85                               \\ \midrule
\reveal         & 81                                    & 29                                     & 59                                  & 38                              & 78                               \\ \midrule
\ivdetect       & 85                                    & 39                                     & 63                                  & 24                              & 83                               \\ \bottomrule
\end{tabular}
\end{table}

%% file: results.tex
\section{Results}
In this section, we present our experimental results with respect to each RQ.\\

\noindent
\textbf{RQ$_1$:} \textit{How do the \deep, \linevul, \reveal, and \ivdetect models perform in a realistic evaluation setting compared to the evaluation setting used in the original studies?}\\

\noindent
\textbf{Approach.}
In this RQ, we evaluate the \deep, the \linevul, the \reveal, and the \ivdetect models using the \dataset testing dataset described in Section \ref{datasets}.
We use the \linevul, \reveal, and the \ivdetect model trained in Section~\ref{sec:replication} for the evaluation, i.e., we train using the same method as the original \linevul, \reveal, and \ivdetect studies, respectively, while using \dataset testing dataset for evaluation.
Note that \linevul, \reveal, and \ivdetect models are trained using real-world vulnerability datasets.
Since the \deep model in Section~\ref{sec:replication} is trained using SARD (an artificially created vulnerability dataset), we train a new \deep model using the \bigvul dataset for evaluation to simulate a fair comparison with the other two models.
First, we generate the \deep model inputs (XFGs) using the \bigvul dataset. 
We obtain a total of 28,294 vulnerable XFGs and 639,047 \nonvulnerable XFGs. Similar to preliminary analysis, we train the \deep model for 50 epochs using the generated XFGs.

To evaluate the trained models (\deep, \linevul, \reveal, and \ivdetect) using the \dataset  testing dataset, we first generate the model inputs (XFG, chunk, CPG, and PDG) for the \dataset testing dataset. These samples are then evaluated using the respective trained models.\\

\noindent
\textbf{Results.}
Table \ref{table:RQ1} shows the results of the performance of the \deep, \linevul, \reveal, and \ivdetect models evaluated using the \dataset testing dataset.
The \deep model obtains an accuracy, precision, recall, F1-score, and AUC of 91\%, 1\%, 87\%, 2\%, and 62\%, respectively.
When comparing these results to those obtained in Section~\ref{sec:replication} (i.e., compared to the results of the \deep model trained and tested using the SARD dataset), we observe a decrease of 7, 86, 11, and 91 percentage points in accuracy, precision, recall, and F1-score, respectively.
\input{tables/original_model_tested_with_realvul}
The \linevul model obtains an accuracy, precision, recall, F1-score, and AUC of 89\%, 1\%, 90\%, 2\%, and 58\%, respectively. In comparison to the results obtained in Section~\ref{sec:replication} (i.e., compared to the results of the \linevul model trained and tested using the \bigvul dataset), we observe a decrease of 7, 95, 88, and 27 percentage points in accuracy, precision, F1-score, and AUC, respectively. However, we also observe an increase of 6 percentage points in the recall.

The \reveal model demonstrates an 89\% accuracy rate, with precision, recall, and F1-score values of 10\%, 80\%, and 17\%, respectively. When comparing these results to the results from Section~\ref{sec:replication} where the \reveal model was trained and tested using the \reveal dataset, we observe a decrease of 19, 21, and 17 percentage points in precision, F1-score, and AUC, respectively, but a 21 percentage points improvement in recall.

The \ivdetect model exhibits an accuracy of 85\%, along with precision, recall, and F1-score metrics of 2\%, 84\%, and 2\% respectively. In comparison to the outcomes from Section~\ref{sec:replication}, where the \ivdetect model underwent training and testing using the \ivdetect dataset, a notable reduction of 37 percentage points in precision, 22 percentage points in the F1-score, and 24 percentage points in AUC can be observed; however, an enhancement of 21 percentage points in recall is also observed.


\begin{figure*}[!tb!]
  \centering
  \begin{subfigure}[t]{0.24\textwidth}
    \centering
    \includegraphics[width=\textwidth]{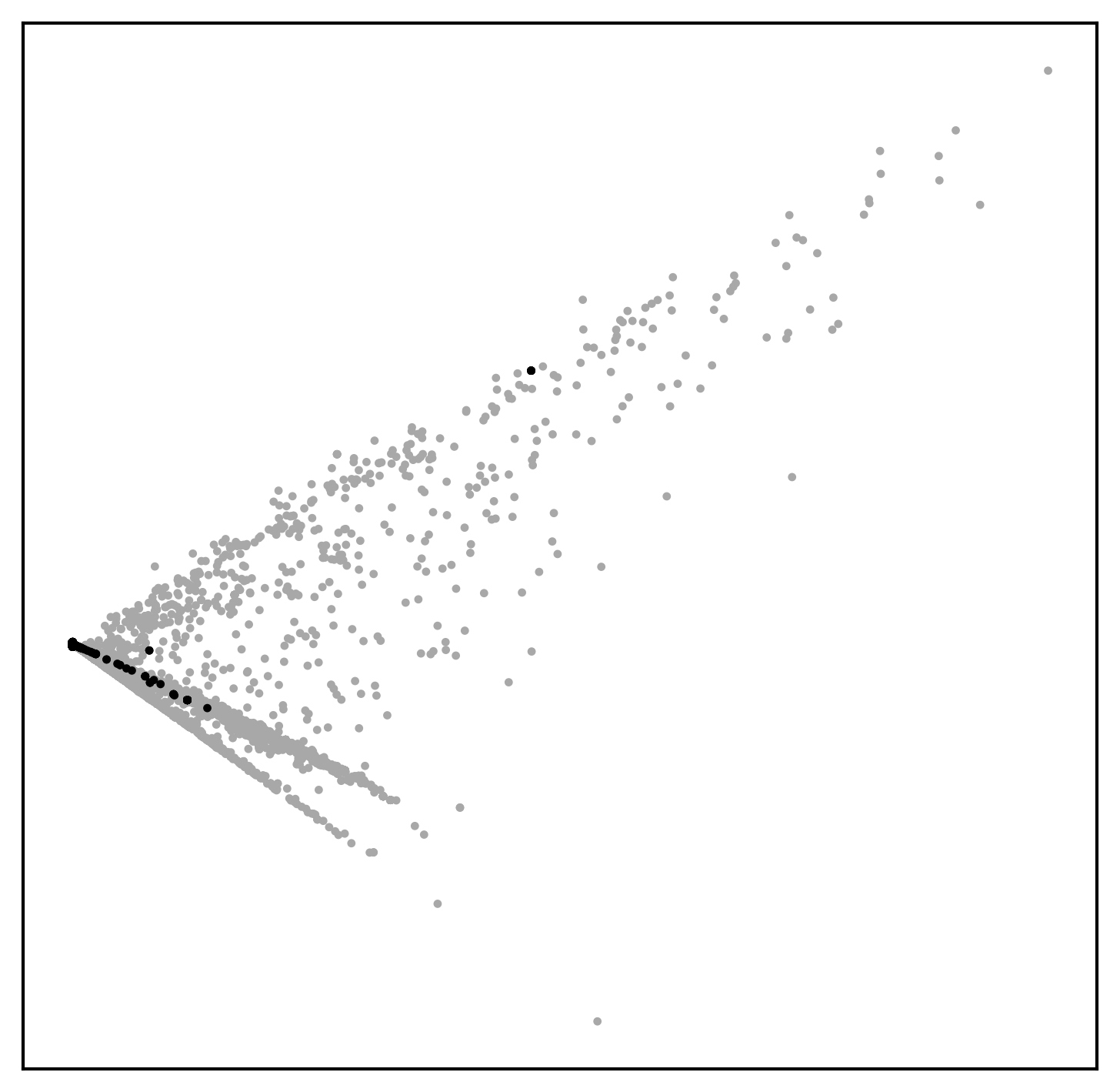}
    \caption{\deep with \dataset.}
    \label{fig:tsneplots4}
  \end{subfigure}
  \begin{subfigure}[t]{0.24\textwidth}
    \centering
    \includegraphics[width=\textwidth]{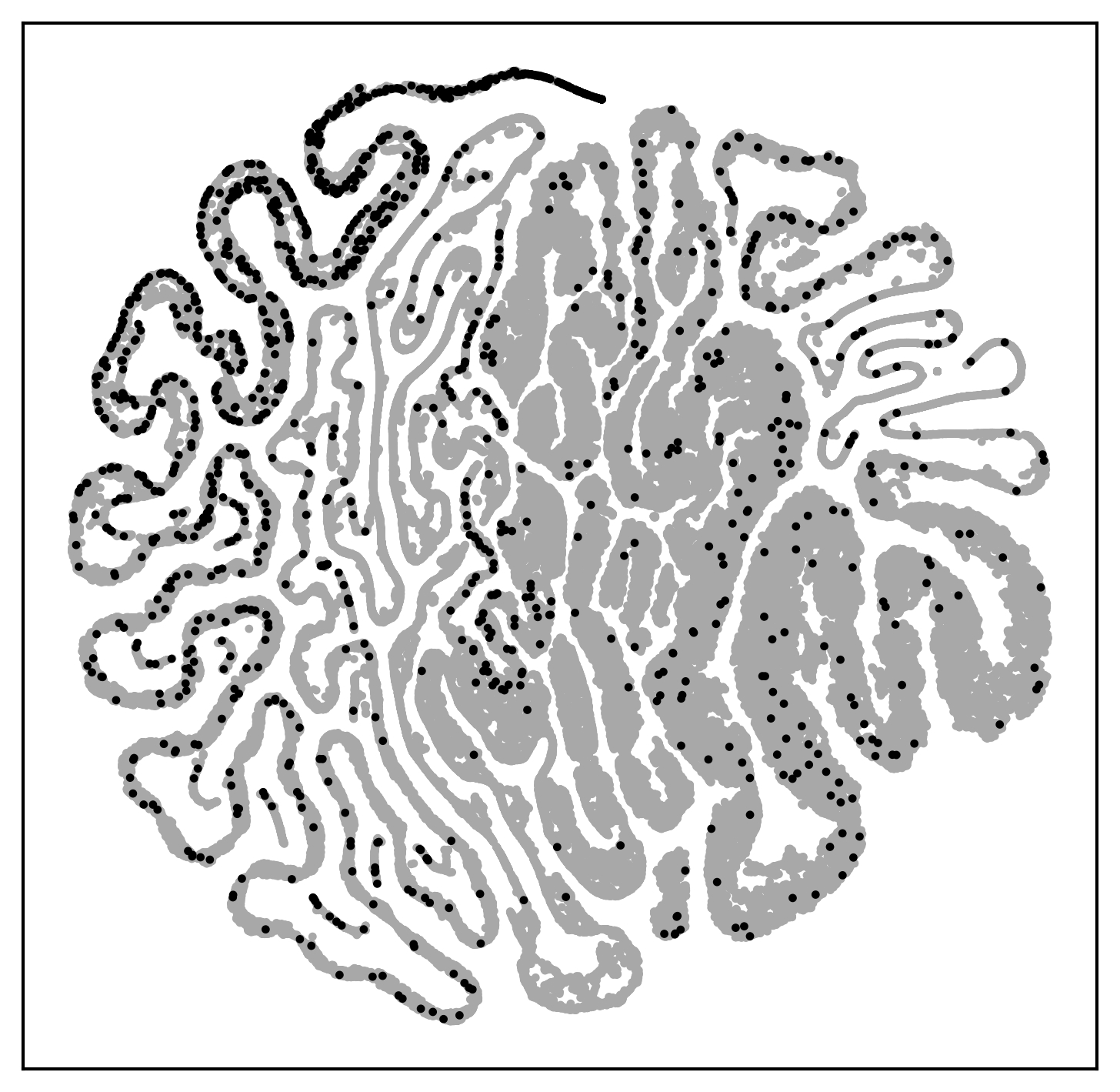}
    \caption{\linevul with \dataset.}
    \label{fig:tsneplots5}
  \end{subfigure}
  \begin{subfigure}[t]{0.233\textwidth}
    \centering
    \includegraphics[width=\textwidth]{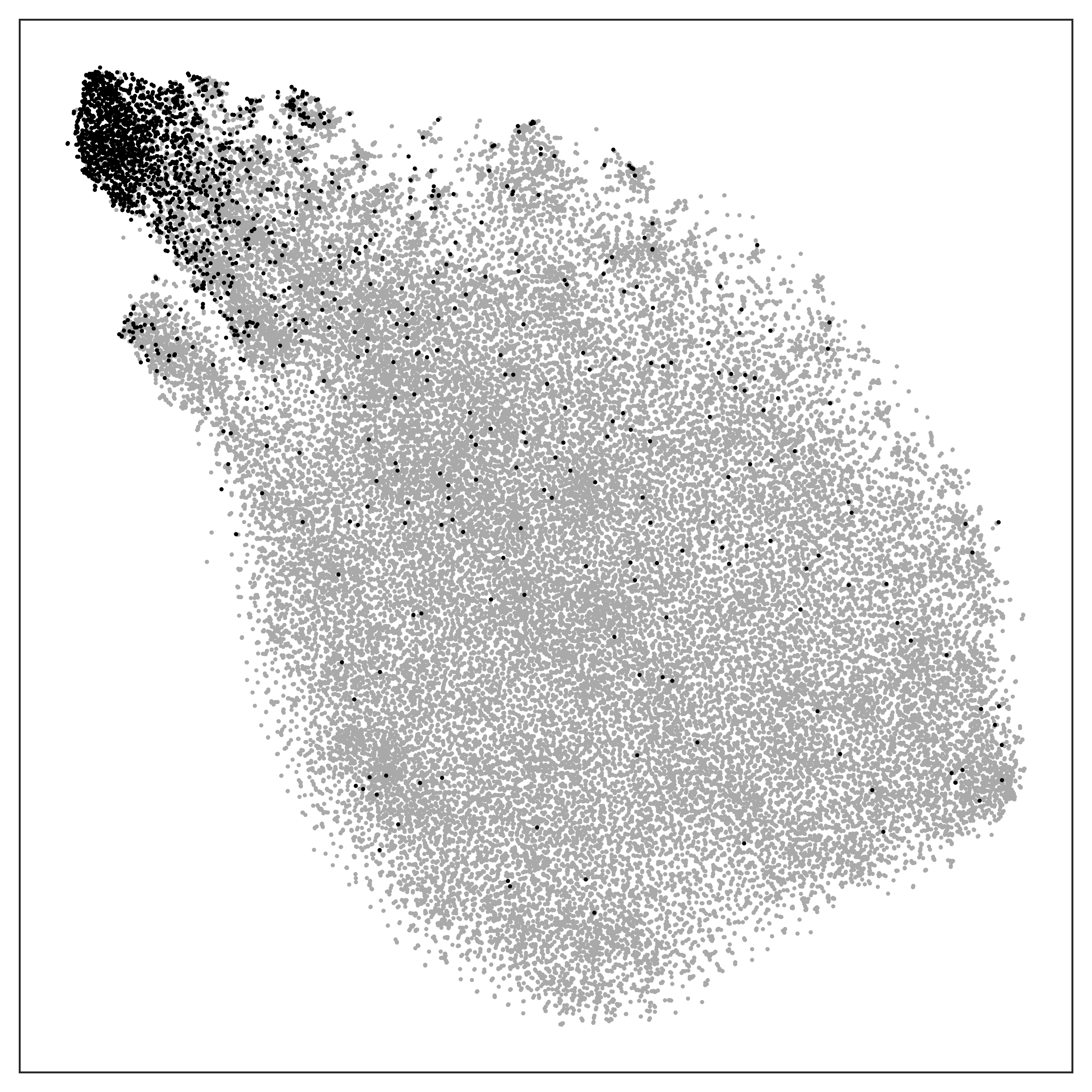}
    \caption{\reveal with \dataset.}
    \label{fig:tsneplots6}
  \end{subfigure}
  \begin{subfigure}[t]{0.233\textwidth}
    \centering
    \includegraphics[width=\textwidth]{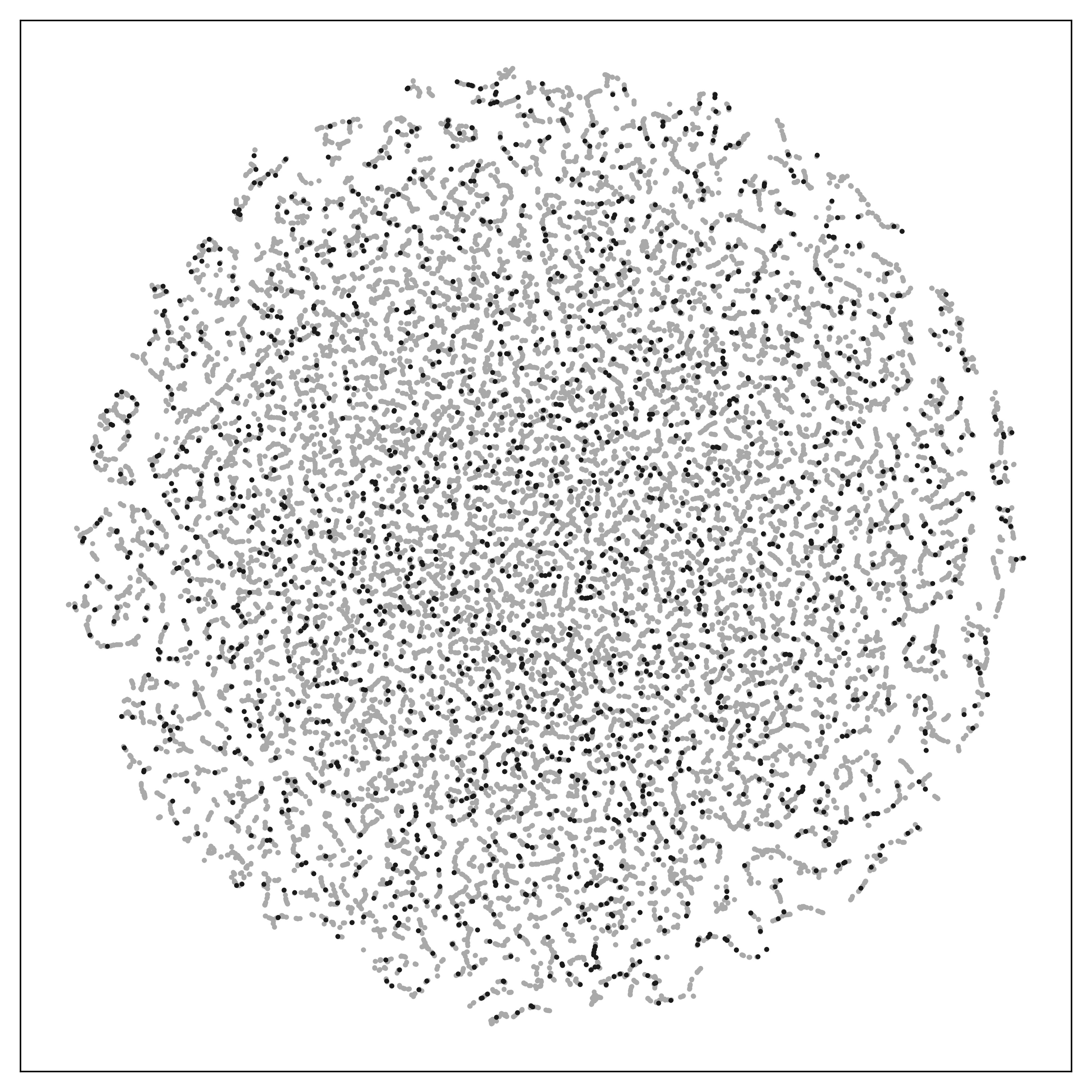}
    \caption{\ivdetect with \dataset.}
    \label{fig:tsneplots7}
  \end{subfigure}
   \caption{Scatter plots showing the class separation between the vulnerable and \nonvulnerable samples. \raisebox{-0.4ex}{\textcolor{black}{\scalebox{2}{\textbullet}}} denotes vulnerable samples and \raisebox{-0.4ex}{\textcolor{gray}{\scalebox{2}{\textbullet}}} denotes \nonvulnerable samples.}
\label{tsneplots}
\end{figure*}
Overall, our findings reveal a substantial decrease in precision for all the models. This suggests that the predictions made by these models contain a considerable number of false positives, which can have a huge impact on the effectiveness of vulnerability detection. Specifically, the \deep model generated 439,494 false positives, while the \linevul, \reveal, and \ivdetect models produced 114,629, 320,128, and 512,284 false positives, respectively. 
Such a large number of false positives may have a large impact on the usability of these models in practice.

To understand the models' performance in terms of class separation, we follow the same approach used in Section~\ref{sec:replication}, and plot embeddings of the models in two-dimensional space. 
It is worth noting that the embeddings are generated for the test split of \dataset dataset.
Figure~\ref{tsneplots} presents the scatter plots of model embeddings.
For all the models, we observe a substantial overlap between vulnerable and \nonvulnerable samples of the \dataset dataset, which indicates that the models do not distinguish between the classes clearly. 
This lack of class separation explains the high number of false positives produced by the \deep, \linevul, \reveal, and \ivdetect models. 

Overall, our findings underscore the importance of evaluating vulnerability detection models using datasets that reflect realistic settings. 
In this regard, our results demonstrate the critical role played by the \dataset dataset in understanding the true capabilities of such models.
By utilizing this dataset, we gain a more accurate understanding of the performance of vulnerability detection models and can identify areas that require improvement.

\boxtext{Existing deep learning-based vulnerability detection models, such as \deep, \linevul, \reveal, and \ivdetect produce a high number of false positives when tested using datasets that represent more accurate real-world testing settings.}

\noindent
\textbf{RQ$_2$:} \textit{How do the \deep, \linevul, \reveal, and \ivdetect models perform in a realistic evaluation setting when trained using a similar realistic training dataset?} \\

\noindent
\textbf{Approach.} 
In RQ$_1$, we use the \dataset dataset for evaluation purposes, i.e., we evaluate the effectiveness of our models using \dataset.
However, in this RQ, we use the \dataset dataset for both model training and evaluation. The training dataset is imbalanced (higher \nonvulnerable samples). Hence, we also investigate the impact of the imbalanced dataset on model performance by training additional models on a balanced dataset. To create a balanced dataset for training, we randomly select \nonvulnerable samples equal to the number of vulnerable samples.



Overall, we train a total of eight models (one imbalanced-trained and one balanced-trained model for \deep, \linevul, \reveal, and \ivdetect techniques) using the same training parameters used in Section~\ref{sec:replication}. 
We evaluate the trained \deep, \linevul, \reveal, and \ivdetect models using the same XFG, chunk, CPG, and PDG test dataset used in RQ$_1$ (i.e., \dataset test dataset).

\noindent \\
\textbf{Results.} Table \ref{table:RQ3_imbalanced}  shows the results of the \deep, \linevul, \reveal, and \ivdetect models that are trained using imbalanced datasets. 
The \deep and \linevul models both exhibit an accuracy of 99\%, with precision, recall, F1-scores at 0\%, and AUC at 50\%. In comparison, the ReVeal model exhibits 95\% accuracy, 2\% precision, 10\% recall, 3\% F1-score, and 50\% AUC. Similarly, the \ivdetect model demonstrates an 81\% accuracy, with precision, recall, F1-score, and AUC recorded at 2\%, 1\%, 0\%, and 49.6\%, respectively. The accuracy is 99\% for both the \deep and \linevul models, while the precision, recall, F1-scores, and AUC are 0\% and 50\%, respectively. 
For the \reveal model, the accuracy, precision, recall, F1-score, and AUC are 95\%, 2\%, 10\%, 3\%, and 50\%, respectively. A common trend across the models is that the precision, recall, and F1-scores are substantially reduced with respect to  the results in Section~\ref{sec:replication} and RQ$_1$.

The high accuracy scores of the models are due to the imbalanced nature of the datasets, which have a larger number of \nonvulnerable samples. This imbalance leads models to often predict samples as \nonvulnerable, boosting accuracy but reducing recall, precision, and F1-scores. The models have difficulty learning from the fewer vulnerable samples because they focus on minimizing a loss function, which discourages wrong predictions. Consequently, they tend to classify samples as \nonvulnerable to reduce errors despite inaccuracies. However, the \reveal model performs better as it uses the SMOTE~\cite{chawla2002smote} algorithm to balance the classes by oversampling the minority (vulnerable samples).

\input{tables/trained_model_with_imbalanced_data}
 
Table \ref{table:RQ3_balanced} shows the performance of the \deep, \linevul, \reveal, and \ivdetect models that are trained using balanced datasets (XFGs, chunks, CPGs, and PDGs).
The results indicate that the \deep model trained on the balanced dataset performs with 89\% accuracy, 1\% precision, 53\% recall, 2\% F1-score, and 51.4\% AUC, which is down by 9, 86, 45, 91, and 36 percentage points, respectively, compared to the \deep model evaluated in Section~\ref{sec:replication}. The \linevul model trained on the balanced dataset performs with 99\% accuracy, 11\% precision, 99\% recall, 20\% F1-score, and 51.4\% AUC. The precision, F1-scores, and AUC are down by 85, 70, and 33.6 percentage points, respectively, compared to the \linevul model evaluated in Section~\ref{sec:replication}. When trained on the balanced dataset, the \reveal model showcases a performance of 91\% accuracy, 31\% precision, 45\% recall, 36\% F1-score, and 51.3\% AUC. In comparison to the \reveal model evaluated in Section~\ref{sec:replication}, the recall, F1-scores, and AUC experience a decline of 14, 2, and 26 percentage points, respectively. The \ivdetect model trained on the balanced dataset performs with 83\% accuracy, 8\% precision, 32\% recall, 6\% F1-score, and 52.3\% AUC. The precision and recall are reduced by 31 percentage points, whereas accuracy, F1-score, and AUC are reduced by 2, 18, and 30 percentage points, respectively, compared to the \ivdetect model of Section~\ref{sec:replication}. \par

Overall, the results suggest that these models still produce a high number of false positives, even when trained using a dataset that is similar to the realistic evaluation dataset. 
\input{tables/trained_model_with_balanced_data}

When we compare how the models did compare to earlier evaluation (RQ$_1$) with the original dataset, we found that the recall scores for the \deep, \reveal, and \ivdetect models dropped by 34, 35, and 52 percentage points, respectively. The differences in how well the \deep, \linevul, \reveal, and \ivdetect models performed could be because of the kind of input they use (like XFG, chunks, CPGs, and PDGs) and model architecture (GNN, CodeBERT). Prior research~\cite{Wan2022} shows that models using graphs usually perform a bit better than those using text, which helps us understand why \reveal (using graphs) performed better than \linevul (using text).

We assess the models when trained on \dataset and tested on their corresponding original datasets (such as SARD, \reveal, BigVul,  and \ivdetect) to cross-check the impact of realistic settings. Our experiments indicate a substantial improvement in performance, with AUC improvements ranging from 8 to 32 percentage points compared to the model trained and tested solely on the \dataset dataset. 
Specifically, on the original test dataset, the \deep model achieved an accuracy of 88\%,  precision of 81\%,  recall of 89\%, F1-score of 84\%, and an AUC of 81\%, observing improvements over its \dataset dataset performance in every metric except accuracy, which decreased by one percentage point.
The \linevul model achieved an accuracy of 80\%, precision of 84\%, recall of 69\%, F1-score of 75\%, and AUC of 79.9\%, showing declines in accuracy and recall by 19 and 30 percentage points, respectively, but gains in precision, F1-score, and AUC.
The \reveal model experienced a 5 percentage point decrease in accuracy while seeing improvements of 4, 6, 5, and 8 percentage points in precision, recall, F1-score, and AUC, respectively.
The \ivdetect model achieved an accuracy of 78\%, precision of 28\%, recall of 48\%, F1-score of 35\%, and AUC of 78.3\%, with a decrease in accuracy by five percentage points but increases in all other metrics.
Overall, the improved performance suggests that the prior datasets might not accurately represent realistic settings, thus overestimating model performance.

The study found that using a balanced dataset for training significantly improves model performance in detecting vulnerabilities. Specifically, the \deep model showed improvements in precision, recall, and F1-scores by 1, 53, and 2 percentage points, respectively, after training on a balanced vs. an imbalanced dataset. This improvement trend was also observed in the \linevul, \reveal, and \ivdetect models. However, despite these gains, the issue of generating false positives remains a challenge for all models.

\boxtext{Despite training the \deep, \linevul, \reveal, and \ivdetect models with a realistic training dataset that closely resembles the realistic test dataset, we observe that these models still generate a high number of false positives (low precision).}

%% file: tables/original_model_tested_with_realvul.tex
\begin{table}[tb]
\caption{Results of the \deep, \linevul, \reveal, and \ivdetect models when evaluated using the \dataset testing dataset.}
\label{table:RQ1}
\centering
\begin{tabular}{@{}crrrrr@{}}
\toprule
\textbf{Model} & \multicolumn{1}{c}{\textbf{\begin{tabular}[c]{@{}c@{}}Accuracy \\ (\%)\end{tabular}}} & \multicolumn{1}{c}{\textbf{\begin{tabular}[c]{@{}c@{}}Precision \\ (\%)\end{tabular}}} & \multicolumn{1}{c}{\textbf{\begin{tabular}[c]{@{}c@{}}Recall \\ (\%)\end{tabular}}} & \multicolumn{1}{c}{\textbf{\begin{tabular}[c]{@{}c@{}}F1\\ (\%)\end{tabular}}} & \multicolumn{1}{c}{\textbf{\begin{tabular}[c]{@{}c@{}}AUC\\ (\%)\end{tabular}}} \\ \midrule 
\deep          & 91                                    & 1                                      & 87                                  & 2                               & 62                               \\ \midrule
\linevul       & 89                                    & 1                                      & 90                                  & 2                               & 58                             \\ \midrule
\reveal        & 89                                    & 10                                     & 80                                  & 17                              & 61                               \\ \midrule
\ivdetect      & 85                                    & 2                                      & 84                                  & 2                               & 59                               \\ \bottomrule
\end{tabular}
\end{table}

%% file: tables/trained_model_with_imbalanced_data.tex
\begin{table}[tb]
\caption{Results of the \deep, \linevul, \reveal,and \ivdetect models when trained using the imbalanced dataset (\dataset dataset).}
\label{table:RQ3_imbalanced}
\centering
\begin{tabular}{@{}crrrrr@{}}
\toprule
\textbf{Model} & \multicolumn{1}{c}{\textbf{\begin{tabular}[c]{@{}c@{}}Accuracy \\ (\%)\end{tabular}}} & \multicolumn{1}{c}{\textbf{\begin{tabular}[c]{@{}c@{}}Precision \\ (\%)\end{tabular}}} & \multicolumn{1}{c}{\textbf{\begin{tabular}[c]{@{}c@{}}Recall \\ (\%)\end{tabular}}} & \multicolumn{1}{c}{\textbf{\begin{tabular}[c]{@{}c@{}}F1\\ (\%)\end{tabular}}} & \multicolumn{1}{c}{\textbf{\begin{tabular}[c]{@{}c@{}}AUC\\ (\%)\end{tabular}}} \\ \midrule
\deep          & 99                                    & 0                                      & 0                                   & 0                               & 50                               \\ \midrule
\linevul       & 99                                    & 0                                      & 0                                   & 0                               & 50                               \\ \midrule
\reveal        & 95                                    & 2                                      & 10                                  & 3                               & 50.1                             \\ \midrule
\ivdetect      & 81                                    & 2                                      & 1                                   & 0                               & 49.6                             \\ \bottomrule
\end{tabular}
\end{table}

%% file: tables/trained_model_with_balanced_data.tex
\begin{table}[tb]
\caption{Results of the \deep, \linevul, \reveal, and \ivdetect models when trained using the balanced dataset (\dataset dataset).}
\label{table:RQ3_balanced}
\centering
\begin{tabular}{@{}crrrrr@{}}
\toprule
\textbf{Model} & \multicolumn{1}{c}{\textbf{\begin{tabular}[c]{@{}c@{}}Accuracy \\ (\%)\end{tabular}}} & \multicolumn{1}{c}{\textbf{\begin{tabular}[c]{@{}c@{}}Precision \\ (\%)\end{tabular}}} & \multicolumn{1}{c}{\textbf{\begin{tabular}[c]{@{}c@{}}Recall \\ (\%)\end{tabular}}} & \multicolumn{1}{c}{\textbf{\begin{tabular}[c]{@{}c@{}}F1\\ (\%)\end{tabular}}} & \multicolumn{1}{c}{\textbf{\begin{tabular}[c]{@{}c@{}}AUC\\ (\%)\end{tabular}}} \\ \midrule
\deep          & 89                                    & 1                                      & 53                                  & 2                               & 51.4                             \\ \midrule
\linevul       & 99                                    & 11                                     & 99                                  & 20                              & 51.4                             \\ \midrule
\reveal        & 91                                    & 31                                     & 45                                  & 36                              & 51.3                             \\ \midrule
\ivdetect      & 83                                    & 8                                      & 32                                  & 6                               & 52.3                             \\ \bottomrule
\end{tabular}
\end{table}

%% file: discussion.tex
\section{Discussion}
In this section, we delve into the false positive predictions and further examine the performance of the models under review by evaluating their effectiveness across different types of vulnerabilities and levels of severity.\\
\textbf{Analysis of false positives.} 
To understand why the studied models produce false positives, we have conducted new analyses from three perspectives.
\begin{enumerate}[leftmargin=1em]
  \begin{figure}[tb]
    \centering
    \includegraphics[width=0.25\textwidth]{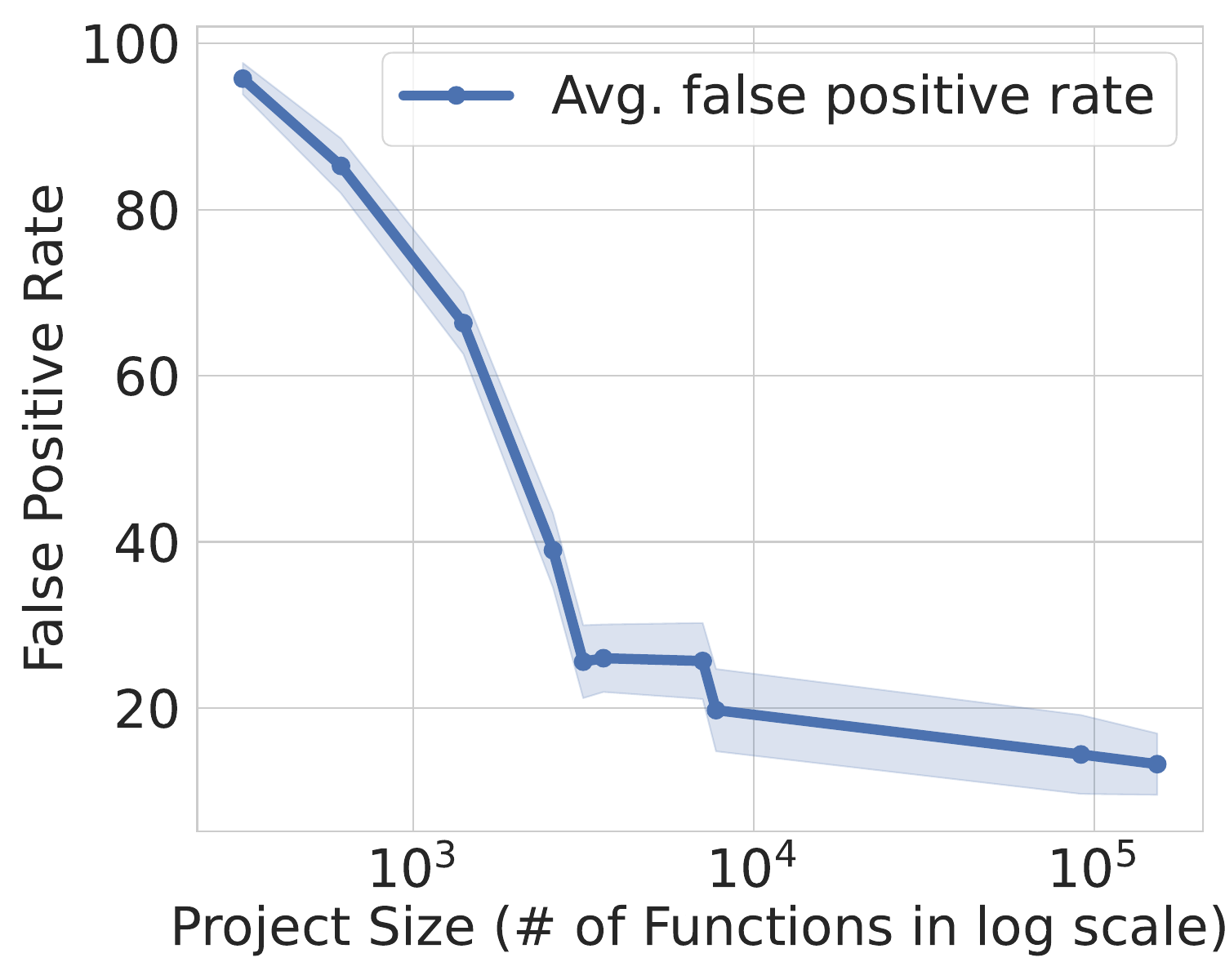}
    \caption{Project Size vs. False Positive Rate for the analyzed projects.}
    \label{fig:fp_vs_size}
  \end{figure}
    \item \textbf{Size of Project.} The motivation for considering project size as a critical factor arises from the notion that larger projects offer a more extensive dataset for model learning. This allows models to better comprehend and identify complex vulnerability patterns, potentially leading to a reduction in false positive rates. In Figure~\ref{fig:fp_vs_size}, we illustrate the influence of project size on false positive rates where we observe a downward trend, i.e., as the size of the project increases, the false positive rate correspondingly decreases. Comparing the smallest project, Jasper, consisting of 183 training and 132 testing samples, with larger projects such as Chrome and Linux, which encompassed 44,303 and 71,706 training samples and 81,351 and 47,181 testing samples, respectively. The findings supported our hypothesis: larger projects exhibited substantially lower false positive rates (85\%-87\% for Chrome and 79\%-85\% for Linux).
    \item \textbf{Code complexity analysis.} Machine learning (ML) models often struggle to comprehend highly complex code~\cite{Xu2022}, which can affect the performance of vulnerability detection tools. To investigate this, we measured the cyclomatic complexity across our test dataset and conducted Mann-Whitney U tests to compare complexities across different categories: true positive vs. false positive ($p=0.043$) and true negative vs. false negative ($p=0.014$). Our comparison suggests that increased complexity leads to model confusion, resulting in higher false positive rates.
    \item \textbf{Overfitting assessment.} ML models may assign excessive importance to specific features, a phenomenon referred to as "overfitting"~\cite{Mehrabi2019}. This tendency can detrimentally impact model performance. For a detailed examination of "overfitting," we employed LIME~\cite{lime}, an explainability tool, to pinpoint the features influencing the models' decisions. An example of the LIME explanation is available in the online Appendix\footref{replication_package}. Note that our analysis encountered challenges with three of our models \deep, \reveal, and \ivdetect—that rely on Joern\footnote{\url{https://github.com/joernio/joern}} to convert source code into abstract graphs. This process introduced complexities in tracing back to the original code for explainability purposes. Consequently, we restricted our analysis to false positives identified solely by the \linevul model. To confirm our findings, we conducted hypothesis testing. First, we create a set of positive tokens from the LIME output (the model predicts positive). Second, we calculate the ratio of positive tokens available in the test dataset samples. Finally, we apply the Mann-Whitney U test with $\alpha=0.05$ to compare the ratio in false positive and true positive samples but find no significant statistical difference ($p=0.13$).
\end{enumerate}

To address this challenge, we propose augmenting the dataset with dead code, a technique that enhances dataset diversity more effectively than simple resampling by creating varied minority class examples, avoiding the reinforcement of misleading patterns~\cite{rebuffi2021}. Dead code, which does not influence program output, introduces variations in textual data through syntactic transformations while preserving the original meaning. By broadening the hypothesis space for deep learning models, augmentation serves as implicit regularization, reducing overfitting to specific tokens and promoting model generalization~\cite{rahman2023}.\\
Influenced by~\cite{na2023}, we augmented the training dataset with dead code to enrich the diversity and mitigate the overemphasis on specific tokens. We employed eleven distinct augmentation strategies, elaborated in the online appendix. We particularly focused on augmenting vulnerable samples to balance them with uncertain (non-vulnerable) samples, thereby enhancing dataset diversity.\\
\input{tables/augmented_performance}
We train the four models in our study using the augmented dataset and evaluate them against the test data. Table~\ref{table: augmeted_performance} provides an overview of the performance metrics. Compared to models that we train using balanced datasets, the AUC substantially improves for all models. Specifically, we find that the text-based model “\linevul” improved more substantially in terms of F1 and AUC than the graph-based models (i.e., \deep, \linevul, \reveal, and \ivdetect). Conversely, we find that the recall of the \linevul model drops by 34 percentage points. The previous model labeled a large proportion of samples as vulnerable and, consequently, achieved an almost perfect recall (at the cost of low precision). \linevul’s improvement in F1 suggests that the improvement in precision (26 percentage points) outweighs the cost of the recall. We conclude that augmentation can mitigate false positives and improve the practical applicability of vulnerability detection models.\par

Note that while text-based models benefit from the variability introduced by dead code, graph-based models may require augmentations that induce structural or data flow changes. Hence, we encourage future work to enhance the effectiveness of dead code injection for graph-based models. The disparity in the impact of dead code injection between text-based and graph-based models stems from their differing code processing approaches. Text-based models may experience enhanced performance due to the variability introduced by dead code, perceiving it as a regularization feature~\cite{Wang2021}. Conversely, graph-based models, which leverage PDGs or DFGs, may regard the dead code as noise, as it does not alter the fundamental program structure or data flow, leading to marginal performance improvements.

\begin{figure*}[tb!]
  \centering
  \begin{subfigure}[t]{0.33\textwidth}
    \centering
    \includegraphics[width=0.9\textwidth]{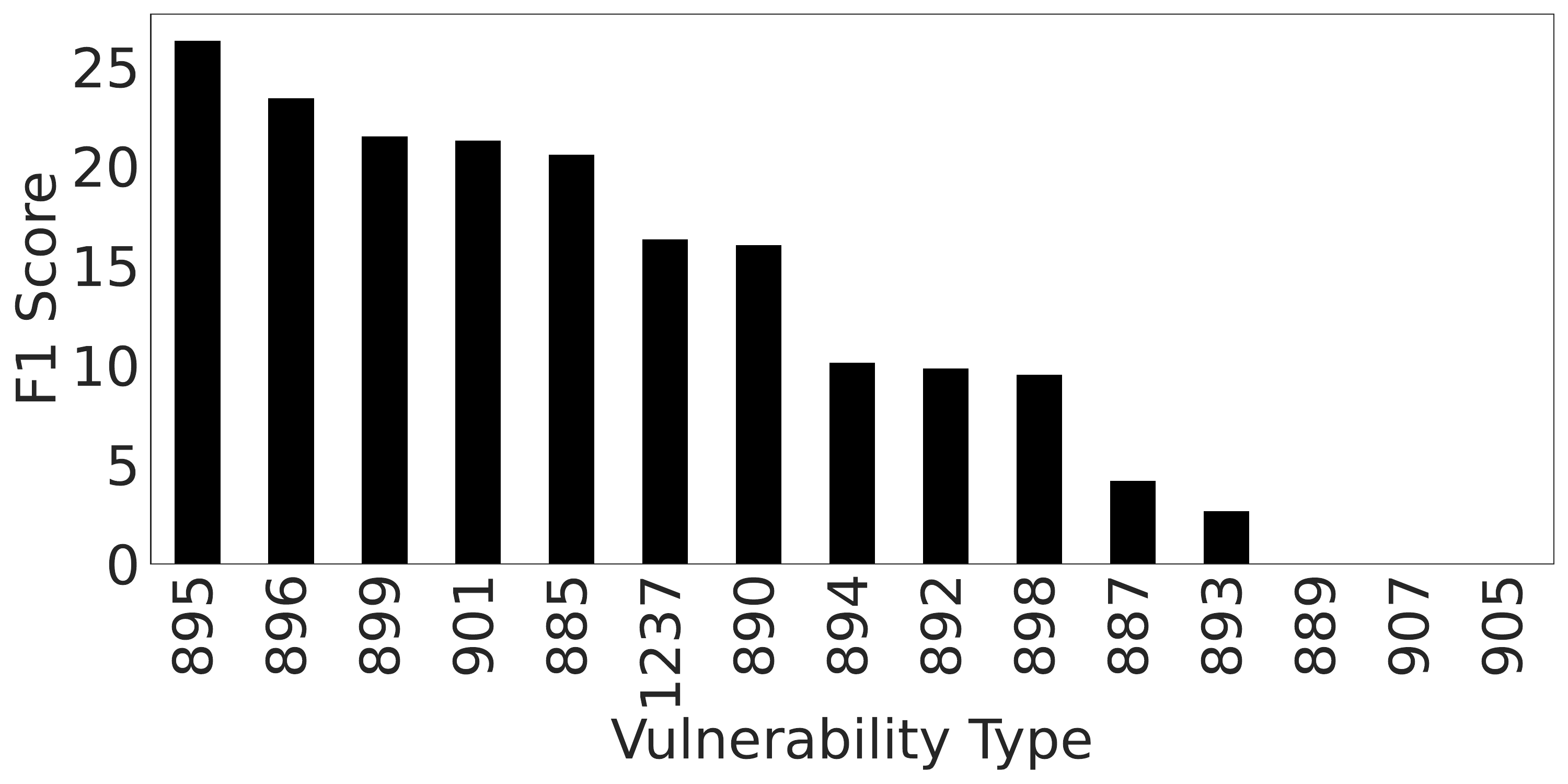}
    \caption{\deep model.}
    \label{fig:f1_cluster_deepwukong}
  \end{subfigure}\quad
  \begin{subfigure}[t]{0.33\textwidth}
    \centering
    \includegraphics[width=0.9\textwidth]{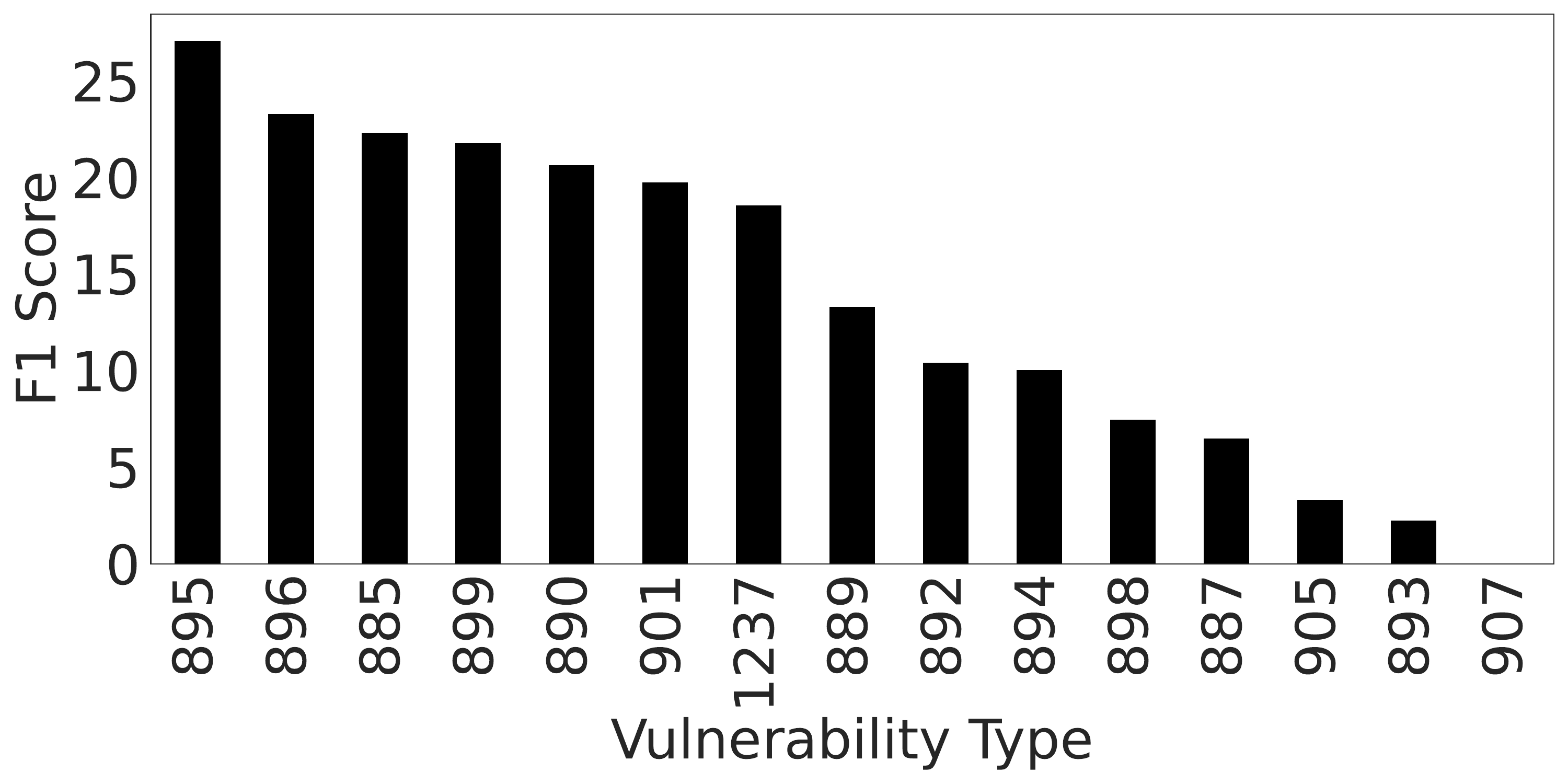}
    \caption{\linevul model.}
    \label{fig:f1_cluster_linevul}
  \end{subfigure}\\
  \begin{subfigure}[t]{0.33\textwidth}
    \centering
    \includegraphics[width=0.9\textwidth]{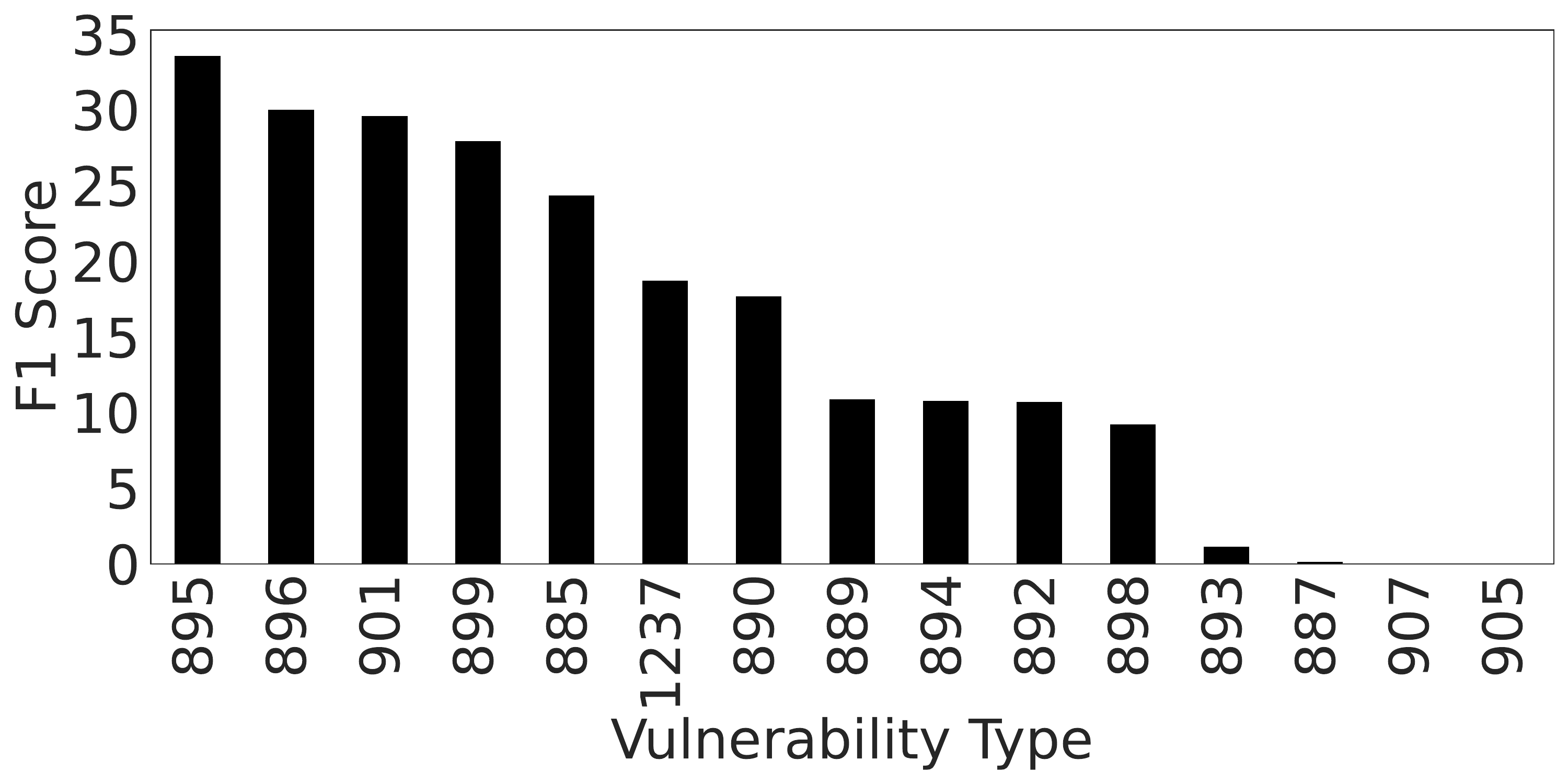}
    \caption{\reveal model.}
    \label{fig:f1_cluster_reveal}
  \end{subfigure}\quad
\begin{subfigure}[t]{0.33\textwidth}
    \centering
    \includegraphics[width=0.9\textwidth]{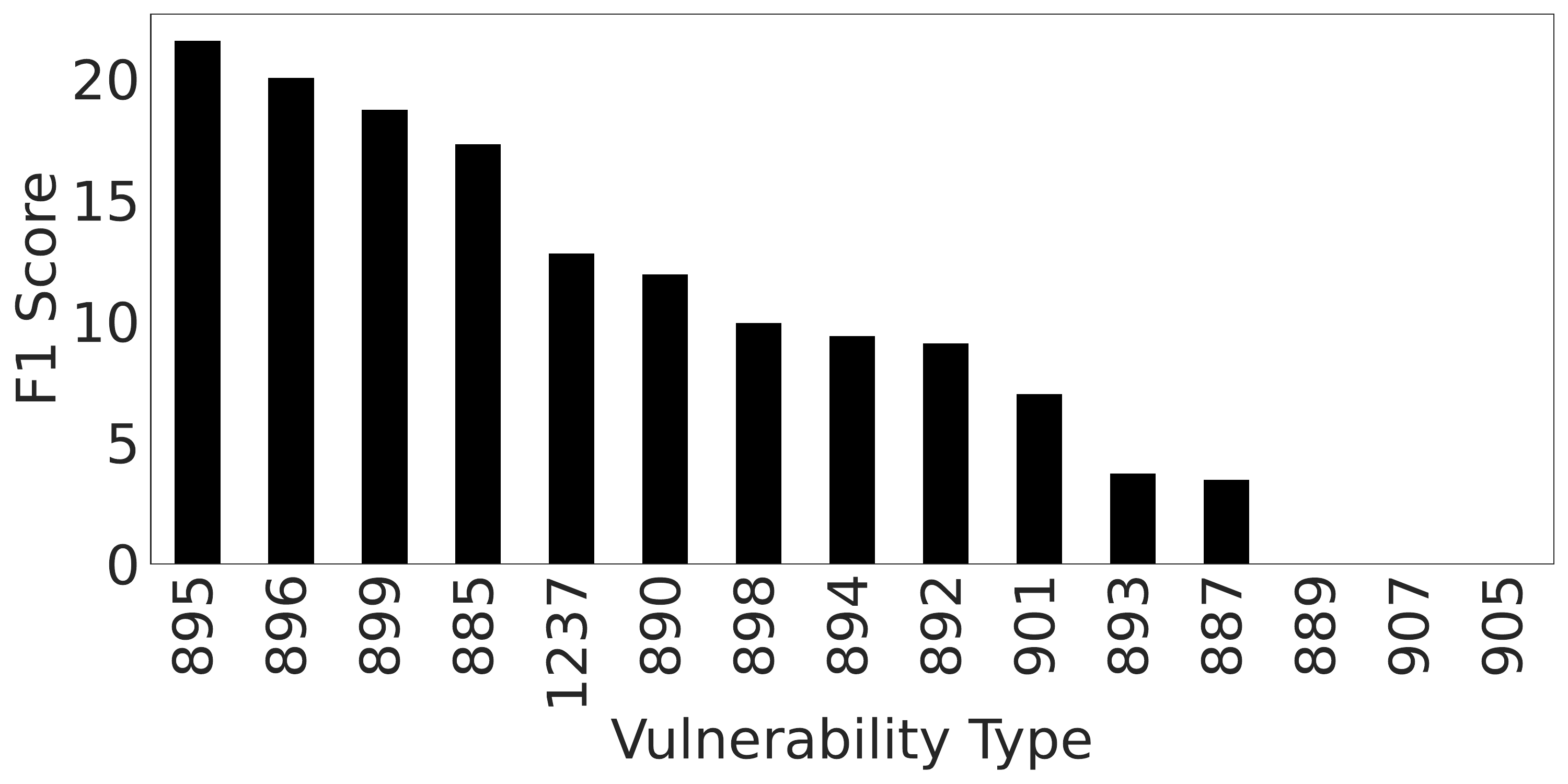}
    \caption{\ivdetect model.}
    \label{fig:f1_cluster_ivdetect}
  \end{subfigure}
   \caption{Performance of the models per SFP cluster.}
\label{f1_cluster}
\end{figure*}

\input{tables/performance_by_severity}
\noindent
\textbf{Performance per vulnerability type.}
In our dataset, each vulnerability sample is associated with a specific Common Weakness Enumeration (CWE). The security community utilized CWEs to group vulnerability types into clusters based on their common characteristics. These clusters are collectively known as the Software Fault Pattern (SFP)~\cite{cwe_cluster}, and they consist of sets of similar CWEs. To assess the performance of the models in detecting vulnerabilities belonging to different clusters, we first calculate the F1-score (using the model trained in $RQ_2$) for each CWE category. Then, we group these CWEs based on the Software Fault Pattern (SFP) clusters.

Figure~\ref{f1_cluster} presents the performance of the studied models in terms of F1-scores across different clusters. Overall, we observe that the performance varies across different clusters, with values up to 26.34\% for \deep, 27.10\% for \linevul, 33.61\% for \reveal, and 21.60\% for \ivdetect. For instance, the models perform exceptionally well in detecting vulnerabilities that belong to clusters 895 or 896, with F1-scores of 33.6\% and 23.31\%, respectively. Cluster 895 includes CWEs associated with information leaks, such as returning a private data structure from a public method (CWE 495) or storing a password in plaintext (CWE 256). On the other hand, cluster 896 comprises CWEs related to tainted input, such as code injection (CWE 94) and improper input validation (CWE 20).

However, the models encounter more difficulties in detecting vulnerabilities belonging to clusters 893, 905, or 907. Cluster 893 contains CWEs related to path traversal (e.g., absolute path traversal CWE 36). Cluster 905 groups vulnerabilities related to predictability-related CWEs (e.g., predictable value range from previous values CWE 343), and cluster 907 consists of miscellaneous CWEs (e.g., client-side enforcement of server-side security CWE 602).

The models performed better at detecting vulnerabilities about information leaks or tainted inputs (clusters 895 and 896), possibly because these issues need less context to detect, with all the needed details present in the code itself. However, identifying vulnerabilities like path traversal (cluster 893) demands understanding the file system's structure, and cluster 905 (predictability issues) might need previous observations. Notably, no model could detect any vulnerabilities in cluster 907, which includes complex issues tied to the compiler, architecture, and design. Vulnerabilities may arise from the interactions between components, which may not be adequately represented by function-level vulnerability datasets. Our findings show that the performance of the model is lower on CWEs related to system-level vulnerabilities (e.g., CWE-893: Improper Path Traversal), suggesting that these models struggle to detect system-level vulnerabilities effectively.


However, it's important to consider that our observation might be influenced by data distribution bias. Typically, machine learning models perform better in specific categories if they have sufficient data from those categories. To verify this assumption, we count the number of samples for each cluster in our dataset and identify the three most and least frequent clusters. We find that clusters 890 (32\%), 896 (25\%), and 892 (14\%) are the most frequent in our dataset, while clusters 905 (0.048\%), 889 (0.024\%), and 907(0.003\%) are the least frequent.

From Figure~\ref{f1_cluster}, we can observe that the models achieve lower performance for the cluster with the highest frequency compared to other clusters with lower frequency. For example, the models did not achieve their highest F1-score when samples belonged to the most frequent cluster in the dataset (cluster 890). Conversely, the models achieve their lowest F1-scores for cluster 907, which is the least frequent sample in our dataset. This indicates that while other factors, such as architecture and input type, influence model performance in detecting vulnerabilities, the distribution of samples also plays a crucial role in making the model performance robust across different SFP clusters (CWE types). \emph{The analysis indicates the inconsistent performance of vulnerability detection models across different CWE clusters, highlighting the influence of data distribution and contextual complexity. Future research in vulnerability detection should concentrate on enhancing data strategies and model architecture to address contextual complexity, as well as incorporating system-level vulnerabilities to improve the overall coverage of vulnerability detection models.}\\

\noindent
\textbf{Performance per vulnerability severity.}
In our investigation, we extend our analysis to consider the severity of vulnerabilities. In our dataset, each vulnerability is categorized into one of three severity types: high, medium, or low. To evaluate the models based on severity, we follow the same approach used for evaluating them per vulnerability type.

Table~\ref{table:severity} presents the performance of the models based on severity. The figure illustrates that all four models achieve relatively higher F1-scores when the severity of a vulnerability is medium, with a slightly lower F1-score for low-severity vulnerabilities. These findings raise concerns about vulnerability detection, as the models demonstrate weaker capabilities in detecting high-severity vulnerabilities.

It is essential to note that this observation may be influenced by the severity distribution within our \dataset dataset. Further analysis revealed that 60.5\% of vulnerabilities in the dataset are assigned medium severity, while only 33.9\% are assigned low severity, and 5\% are assigned high severity.
This distribution of severity levels could be contributing to model performances across different severity types. \emph{The investigation reveals models are more effective at detecting medium-severity vulnerabilities, with decreased accuracy for high-severity ones. This underscores the need for future research to focus on enhancing the detection of high-severity vulnerabilities, considering the potential impact of severity distribution in training datasets.}\\
\input{tables/llm_performance}
\noindent
\textbf{Efficacy of Large Language Models.} Recent advancements in generative AI have led to the development of specialized Large Language Models (LLMs) for natural language understanding and code interpretation, such as CodeLlama~\cite{rozière2024code} and Mixtral~\cite{jiang2024mixtral}. CodeLlama, a 7 B parameter model, excels in parsing and generating code across multiple programming languages, trained on a vast array of sources, including open-source projects and technical documentation, showcasing capabilities in code completion, bug detection, and retrieval.
Mixtral~\cite{jiang2024mixtral} (7 B parameter) is a Sparse Mixture-of-Experts (SMoE) network. Its architecture can be divided into three parts: an input embedding layer, several decoder blocks, and a language model decoding head. The model is trained using supervised fine-tuning on an instruction dataset and Direct Preference Optimization (DPO) on a paired feedback dataset. The pre-training involves data sourced from the open web. In terms of performance, Mixtral has shown impressive results in various benchmarks, surpassing Llama 2 70B and equating or outperforming Open AI GPT-3.5 in most standard tests despite not being specifically tailored for code tasks.\par
We fine-tuned these models using parameter-efficient fine-tuning~\cite{dettmers2023qlora} to detect vulnerabilities in C++ functions, using the \dataset dataset for training and a separate test dataset for evaluation. The results, detailed in Table~\ref{table: llm_performance}, indicate that these LLMs outperform other tools like \deep, \reveal, and \ivdetect in terms of AUC. However, when compared to \linevul, CodeLlama and Mixtral's AUC lags by 14.8 and 10.8 percentage points, respectively. \par
The lower performance of the LLMs in comparison to \linevul may stem from multiple reasons, such as data efficiency, model generalizability, and the trade-offs inherent in model complexity. LLMs, with their extensive parameter count, may not always generalize well to specific tasks outside their training scope, leading to suboptimal performance. In contrast, models like \linevul, through targeted training and optimization, can achieve better alignment with specific tasks. A similar observation has also been observed in another task where a RoBERTa model outperforms Llama 2~\cite{touvron2023llama} and the Mixtral model~\cite{Iraqi_2023}. \emph{The study shows that specialized tools (\linevul) outperform general large language models (LLMs) in vulnerability detection, highlighting the need for future work to improve LLMs' specificity and efficiency in vulnerability detection tasks.}

%% file: tables/augmented_performance.tex
\setlength{\tabcolsep}{4pt}
\begin{table}[tb]
\caption{Results of the \deep, \linevul, \reveal, and \ivdetect models when trained using the augmented data. (\textit{Values in parentheses show the difference from training on a non-augmented balanced dataset})}
\label{table: augmeted_performance}
\centering
\begin{tabular}{@{}p{1.4cm}ccccc@{}}
\toprule
\textbf{Model} & \multicolumn{1}{c}{\textbf{\begin{tabular}[c]{@{}c@{}}Accuracy \\ (\%)\end{tabular}}} & \multicolumn{1}{c}{\textbf{\begin{tabular}[c]{@{}c@{}}Precision \\ (\%)\end{tabular}}} & \multicolumn{1}{c}{\textbf{\begin{tabular}[c]{@{}c@{}}Recall \\ (\%)\end{tabular}}} & \multicolumn{1}{c}{\textbf{\begin{tabular}[c]{@{}c@{}}F1\\ (\%)\end{tabular}}} & \multicolumn{1}{c}{\textbf{\begin{tabular}[c]{@{}c@{}}AUC\\ (\%)\end{tabular}}} \\ \midrule
\deep          & 91 ($\uparrow$ 2)                                    & 4 ($\uparrow$ 3)                                      & 48 ($\downarrow$ 4)                                  & 7 ($\uparrow$ 5)                               & 58 ($\uparrow$ 6.6)                               \\ \midrule
\linevul       & 98 ($\downarrow$ 1)                                    & 37 ($\uparrow$ 26)                                     & 65 ($\downarrow$ 34)                                  & 46 ($\uparrow$ 26)                              & 81.8 ($\uparrow$ 30.4)                             \\ \midrule
\reveal        & 92 ($\downarrow$ 1)                                    & 33 ($\uparrow$ 2)                                     & 47 ($\uparrow$ 2)                                  & 38 ($\uparrow$ 2)                              & 57 ($\uparrow$ 5.6)                               \\ \midrule
\ivdetect      & 78 ($\downarrow$ 5)                                    & 10 ($\uparrow$ 2)                                     & 35 ($\uparrow$ 3)                                  & 16 ($\uparrow$ 10)                              & 53 ($\uparrow$ 0.7)                               \\ \bottomrule
\end{tabular}
\end{table}

%% file: tables/performance_by_severity.tex
\begin{table}[tb]
\caption{Performance of the models per severity.}
\label{table:severity}
\centering
\begin{tabular}{@{}crrr@{}}
\toprule
\textbf{Model} & \multicolumn{1}{c}{\textbf{Low (\%)}} & \multicolumn{1}{c}{\textbf{Medium (\%)}} & \multicolumn{1}{c}{\textbf{High (\%)}} \\ \midrule
\deep          & 18.26                            & 20.56                               & 9.7                               \\ \midrule
\linevul       & 20.11                            & 21.28                               & 10.89                             \\ \midrule
\reveal        & 21.66                            & 26.57                               & 10.07                             \\ \midrule
\ivdetect      & 15.08                            & 17.46                               & 9.06                              \\ \bottomrule
\end{tabular}
\end{table}

%% file: tables/llm_performance.tex
\begin{table}[tb]
\caption{Results of the CodeLlama and Mixtral models when trained using the augmented data.}
\label{table: llm_performance}
\centering
\begin{tabular}{@{}crrrrr@{}}
\toprule
\textbf{Model} & \multicolumn{1}{c}{\textbf{\begin{tabular}[c]{@{}c@{}}Accuracy \\ (\%)\end{tabular}}} & \multicolumn{1}{c}{\textbf{\begin{tabular}[c]{@{}c@{}}Precision \\ (\%)\end{tabular}}} & \multicolumn{1}{c}{\textbf{\begin{tabular}[c]{@{}c@{}}Recall \\ (\%)\end{tabular}}} & \multicolumn{1}{c}{\textbf{\begin{tabular}[c]{@{}c@{}}F1\\ (\%)\end{tabular}}} & \multicolumn{1}{c}{\textbf{\begin{tabular}[c]{@{}c@{}}AUC\\ (\%)\end{tabular}}} \\ \midrule
CodeLlama      & 80                                    & 34                                      & 54                                  & 42                               & 67                               \\ \midrule
Mixtral        & 83                                    & 38                                      & 59                                  & 46                               & 71                               \\ \bottomrule
\end{tabular}
\end{table}

%% file: related_works.tex
\section{Related Work}
In this section, we discuss the related works and reflect
on how they compare with ours.
\vspace{2mm}

\noindent
\textbf{Vulnerability detection datasets}. Grahn et al.~\cite{grahn2021analysis} found that some of the vulnerability detection datasets~\cite{russell2018automated, fan2020ac} were not very useful for training models. To address this, they proposed a new dataset called \textit{Wild C}, comprising 10.3 million C/C++ files from multiple open-source projects. However, the drawback of \textit{Wild C} is that it lacks labels for each file, making it unsuitable for building vulnerability detection classification models. In contrast, the \dataset dataset offers the complete source code of projects along with labeled samples as vulnerable or \nonvulnerable, making it more suitable for training vulnerability detection models.


The \bigvul dataset~\cite{fan2020ac} is a collection of C/C++ functions from 348 open-source GitHub projects, used in vulnerability detection studies. However, it lacks representation of the entire codebase since it only includes functions from vulnerability-fixing commits. To address this, we introduce the \dataset dataset, which includes all source code from the top ten real-world projects by vulnerability counts in \bigvul. This provides a more comprehensive and realistic dataset for training and evaluating vulnerability detection models.


\vspace{2mm}
\noindent
\textbf{Studies on vulnerability detection techniques.} 
Several studies explored the effectiveness of traditional machine learning techniques~\cite{neuhaus2009beauty, yan2017machine,zheng2020impact}.
Neuhaus et al.~\cite{neuhaus2009beauty} investigated the prevalence of software vulnerabilities in Red Hat packages using Support Vector Machines (SVM) \cite{hearst1998support}.  The study analyzed the defect data from over 3,241 Red Hat packages and evaluated the effectiveness of SVM in identifying vulnerabilities in software packages. Zheng et al.~\cite{zheng2020impact} examined the effectiveness of different machine learning techniques, like Decision Tree \cite{quinlan1996learning}, Random Forests \cite{breiman2001random}, k-nearest neighbors (KNN) \cite{peterson2009k}, and SVM, in detecting software vulnerabilities. The results of the study provided insights into the strengths and weaknesses of different machine learning techniques for vulnerability detection. Yan et al.~\cite{yan2017machine} employed a combination of static analysis, machine learning, and typestate modeling techniques for static detection of use-after-free vulnerabilities in software. Lomio et al.~\cite{lomio2022just} investigated whether machine learning algorithms like SVM, KNN, Decision Tree, and Boosting algorithms \cite{freund1996experiments, friedman2001greedy, chen2016xgboost} could improve the performance of Just-in-Time software vulnerability detection utilizing various software metrics (process metrics, product metrics, and text metrics). 

Other studies focused on exploring the potential of various deep learning methods to detect vulnerabilities~\cite{li2017software, li2018vuldeepecker, cheng2019static, fu2022linevul, zhou2019devign, cheng2021deepwukong, li2021sysevr}.
For example, Convolutional Neural Networks (CNN) \cite{lecun1995convolutional} have been used to forecast software defects and locate defective source code \cite{li2017software}. Li et al.~\cite{li2021sysevr} made multiple kinds of deep neural networks such as CNN, LSTM \cite{hochreiter1997long}, and GRU \cite{chung2014empirical} to detect vulnerabilities.  
Vuldeepecker \cite{li2018vuldeepecker} detects resource management issues and buffer overflows by training an LSTM model with code embedding and data-flow information of a program. 
VGDetector \cite{cheng2019static} uses a control flow graph and a graph convolutional network \cite{kipf2016semi} to detect control-flow vulnerabilities. Zhou et al.~\cite{zhou2019devign} pinpointed bugs at the method level using Graph Neural Networks (GNN) and program dependence graph. 

The aforementioned work used either synthetic datasets (e.g., SARD), datasets created using oracles like static analysis tools (e.g., \cite{scandariato2014predicting, zheng2021d2a}), or real-world datasets that do not accurately reflect the realistic settings (e.g., \cite{zhou2019devign, fan2020ac, chakraborty2021deep}).
This motivated us to create a new dataset that tackles the lack of realistic evaluation settings in the existing techniques. 
 We create \dataset dataset, which is more comprehensive than the existing vulnerability detection datasets.

Chakraborty et al.~\cite{chakraborty2021deep} introduced the \reveal dataset, similar to our study, to highlight limitations in existing deep learning-based vulnerability detection models. However, they have considered only unchanged functions of vulnerability fixing commits as the \nonvulnerable sample. Additionally, the dataset suffers from label inconsistency. Fu et al.~\cite{fu2022linevul} proposed \linevul, a CodeBERT-based model \cite{feng2020codebert}, for vulnerability detection. Our study differs from the prior work since we propose a realistic dataset \dataset which is free from the limitations of prior studies. Additionally, we evaluate four \textit{state-of-the-art} techniques (i.e., \cite{cheng2021deepwukong}, \cite{fu2022linevul}, \cite{chakraborty2021deep} ) on \dataset, providing the empirical evidence that the machine learning models exhibit limited performance when assessed in real-world settings. Ding et al.~\cite{ding2024} share similar insights regarding data quality in vulnerability datasets; however, they maintained data quality through heuristic-based manual verification and by using a vulnerability database. In contrast, our study aims to create a more realistic dataset by maintaining the ratio of vulnerable to uncertain functions and reducing label inconsistency. Moreover, we found that augmenting the dataset can improve model performance.

%% file: threats.tex
\section{Threats to Validity}

Our study uses the \bigvul dataset \cite{fan2020ac} to construct \dataset.
It is possible that some vulnerable samples in the \bigvul dataset are mislabelled. 
However, the labeled samples in the dataset were manually verified by Fan et al.~\cite{fan2020ac}.
Also, the \dataset dataset may not fully represent all real-world scenarios since it is constructed using only ten open-source projects. 
However, it is worth noting that these ten projects are well-established and popular (e.g., Chrome~\footnote{\url{https://chromium.googlesource.com/chromium/src/}}and Linux~\footnote{\url{https://github.com/torvalds/linux}} and have been extensively studied and utilized in previous research~\cite{zhou2019devign, chakraborty2021deep}. 
In \dataset dataset, we labeled all the unchanged functions of a repository as \nonvulnerable. It is possible that undiscovered vulnerabilities may exist in these functions. However, we are limited by the lack of the knowledge required to discover undiscovered vulnerabilities and the threat is present in prior datasets as well~\cite{fan2020ac, zhou2019devign, chakraborty2021deep}.
In RQ$_2$, we use balanced datasets where the number of vulnerable and \nonvulnerable samples is equal. However, the random selection of samples for the balanced datasets may impact our results since different random selections can lead to different findings \cite{hagan1997neural}. 
To address this issue, it is recommended to train the models multiple times with various sample sets and examine the outcomes. Unfortunately, this approach was not feasible due to our constrained computational resources.

In our study, we excluded methods that have identical MD5 hashes but conflicting labels (label inconsistency). Despite this, our dataset might still have subtle label inconsistencies that arise from variations in whitespace or new lines, which are challenging to detect. To mitigate this, we employed a regex-based strategy to merge multiple instances of new lines and whitespace into single occurrences. Nonetheless, there may be scenarios where this method falls short.

Furthermore, we focus our work on assessing the efficacy of deep learning-based techniques, which can limit the generalizability of our findings to other techniques.
Future works should consider evaluating other techniques like static and dynamic analysis tools on the \dataset dataset and analyze the results.  
%




%% file: implication.tex
\section{Implication}
Below, we distill the implication of our findings 
for the development of research communities.

\noindent The \dataset dataset can be used to develop tools to assist developers in mitigating vulnerabilities. For example,
\begin{itemize}[leftmargin=1em]

    \item \textbf{Facilitating research studies for proposing risk assessment and prioritization techniques.} The dataset contains severity information for each vulnerable sample. This data can be used to train a model to predict the presence of vulnerabilities and assess the potential risk or severity. This aids in prioritizing security efforts and resource allocation.
    
    \item \textbf{Supporting research on automated patch generation.} The dataset contains before-fix and after-fix versions of vulnerable functions. This feature can inform the development of automated patch-generation tools.
    \item \textbf{Characterizing the evolution of software with respect to its vulnerabilities.} The dataset contains vulnerability data of 10 common and popular projects that span ten years of history. It contains 270,919 samples, of which 5,528 are vulnerable. Future research could explore whether severe vulnerabilities have become more or less prevalent over time or how coding practices have evolved to mitigate vulnerabilities in these projects.
\end{itemize}


%% file: conclusion.tex
\section{Conclusion}
In this paper, we study the performance of deep learning-based vulnerability detection models in realistic vulnerability detection settings.
First, we create a new comprehensive, realistic vulnerability detection dataset, called \dataset.
\dataset contains complete source code samples of ten diverse real-world open-source projects.
Then, we evaluate four \textit{state-of-the-art} models, \linevul, \deep, \reveal, and \ivdetect on the ~\dataset dataset.
Our evaluation indicates a considerable decrease in model performance, as evidenced by a drop in precision and F1 scores of up to 95 and 91 percentage points, respectively. Our investigation also reveals that the embeddings generated by these models depict a substantial overlap between vulnerable and \nonvulnerable samples. 
This suggests that such models struggle to differentiate between vulnerable and \nonvulnerable samples in the ~\dataset dataset, resulting in a high number of false positives. 
Finally, we observe fluctuations in model performance based on vulnerability characteristics (e.g., vulnerability types and severity).
Our study argues that when it comes to
identifying vulnerabilities in realistic vulnerability detection settings, things may not be as good as they seem, and there is a need for improved model design and evaluation approaches to achieve more accurate vulnerability detection performance.

%% file: main.bbl
\begin{thebibliography}{10}
\providecommand{\url}[1]{#1}
\csname url@samestyle\endcsname
\providecommand{\newblock}{\relax}
\providecommand{\bibinfo}[2]{#2}
\providecommand{\BIBentrySTDinterwordspacing}{\spaceskip=0pt\relax}
\providecommand{\BIBentryALTinterwordstretchfactor}{4}
\providecommand{\BIBentryALTinterwordspacing}{\spaceskip=\fontdimen2\font plus
\BIBentryALTinterwordstretchfactor\fontdimen3\font minus
  \fontdimen4\font\relax}
\providecommand{\BIBforeignlanguage}[2]{{%
\expandafter\ifx\csname l@#1\endcsname\relax
\typeout{** WARNING: IEEEtran.bst: No hyphenation pattern has been}%
\typeout{** loaded for the language `#1'. Using the pattern for}%
\typeout{** the default language instead.}%
\else
\language=\csname l@#1\endcsname
\fi
#2}}
\providecommand{\BIBdecl}{\relax}
\BIBdecl

\bibitem{chakraborty2021deep}
S.~Chakraborty, R.~Krishna, Y.~Ding, and B.~Ray, ``Deep learning based
  vulnerability detection: Are we there yet,'' \emph{IEEE Transactions on
  Software Engineering}, 2021.

\bibitem{chakraborty2023rlocator}
P.~Chakraborty, M.~Alfadel, and M.~Nagappan, ``Rlocator: Reinforcement learning
  for bug localization,'' \emph{arXiv preprint arXiv:2305.05586}, 2023.

\bibitem{SARD_link}
N.~I. of~Standards and Technology, ``{N}{I}{S}{T} {S}oftware {A}ssurance
  {R}eference {D}ataset --- samate.nist.gov,''
  \url{https://samate.nist.gov/SARD}, [Accessed 28-07-2023].

\bibitem{cheng2021deepwukong}
X.~Cheng, H.~Wang, J.~Hua, G.~Xu, and Y.~Sui, ``Deepwukong: Statically
  detecting software vulnerabilities using deep graph neural network,''
  \emph{ACM Transactions on Software Engineering and Methodology (TOSEM)},
  vol.~30, no.~3, pp. 1--33, 2021.

\bibitem{zheng2020impact}
W.~Zheng, J.~Gao, X.~Wu, F.~Liu, Y.~Xun, G.~Liu, and X.~Chen, ``The impact
  factors on the performance of machine learning-based vulnerability detection:
  A comparative study,'' \emph{Journal of Systems and Software}, vol. 168, p.
  110659, 2020.

\bibitem{lin2020software}
G.~Lin, S.~Wen, Q.-L. Han, J.~Zhang, and Y.~Xiang, ``Software vulnerability
  detection using deep neural networks: a survey,'' \emph{Proceedings of the
  IEEE}, vol. 108, no.~10, pp. 1825--1848, 2020.

\bibitem{fan2020ac}
J.~Fan, Y.~Li, S.~Wang, and T.~N. Nguyen, ``Ac/c++ code vulnerability dataset
  with code changes and cve summaries,'' in \emph{Proceedings of the 17th
  International Conference on Mining Software Repositories}, 2020, pp.
  508--512.

\bibitem{fu2022linevul}
M.~Fu and C.~Tantithamthavorn, ``{LineVul},'' in \emph{Proceedings of the 19th
  International Conference on Mining Software Repositories}.\hskip 1em plus
  0.5em minus 0.4em\relax {ACM}, May 2022.

\bibitem{Li2021}
Y.~Li, S.~Wang, and T.~N. Nguyen, ``Vulnerability detection with fine-grained
  interpretations,'' in \emph{Proceedings of the 29th ACM Joint Meeting on
  European Software Engineering Conference and Symposium on the Foundations of
  Software Engineering}, ser. ESEC/FSE ’21.\hskip 1em plus 0.5em minus
  0.4em\relax ACM, Aug. 2021.

\bibitem{van2008visualizing}
L.~Van~der Maaten and G.~Hinton, ``Visualizing data using t-sne.''
  \emph{Journal of machine learning research}, vol.~9, no.~11, 2008.

\bibitem{zhou2019devign}
Y.~Zhou, S.~Liu, J.~Siow, X.~Du, and Y.~Liu, ``Devign: Effective vulnerability
  identification by learning comprehensive program semantics via graph neural
  networks,'' \emph{Advances in neural information processing systems},
  vol.~32, 2019.

\bibitem{zheng2021d2a}
Y.~Zheng, S.~Pujar, B.~Lewis, L.~Buratti, E.~Epstein, B.~Yang, J.~Laredo,
  A.~Morari, and Z.~Su, ``D2a: a dataset built for ai-based vulnerability
  detection methods using differential analysis,'' in \emph{2021 IEEE/ACM 43rd
  International Conference on Software Engineering: Software Engineering in
  Practice (ICSE-SEIP)}.\hskip 1em plus 0.5em minus 0.4em\relax IEEE, 2021, pp.
  111--120.

\bibitem{li2018vuldeepecker}
Z.~Li, D.~Zou, S.~Xu, X.~Ou, H.~Jin, S.~Wang, Z.~Deng, and Y.~Zhong,
  ``Vuldeepecker: A deep learning-based system for vulnerability detection,''
  \emph{arXiv preprint arXiv:1801.01681}, 2018.

\bibitem{Walden2020}
J.~Walden, ``The impact of a major security event on an open source project,''
  in \emph{Proceedings of the 17th International Conference on Mining Software
  Repositories}.\hskip 1em plus 0.5em minus 0.4em\relax {ACM}, Jun. 2020.

\bibitem{feng2020codebert}
Z.~Feng, D.~Guo, D.~Tang, N.~Duan, X.~Feng, M.~Gong, L.~Shou, B.~Qin, T.~Liu,
  D.~Jiang \emph{et~al.}, ``Codebert: A pre-trained model for programming and
  natural languages,'' \emph{arXiv preprint arXiv:2002.08155}, 2020.

\bibitem{cangea2018towards}
C.~Cangea, P.~Veli{\v{c}}kovi{\'c}, N.~Jovanovi{\'c}, T.~Kipf, and P.~Li{\`o},
  ``Towards sparse hierarchical graph classifiers,'' \emph{arXiv preprint
  arXiv:1811.01287}, 2018.

\bibitem{Joyce2011}
J.~M. Joyce, \emph{Kullback-Leibler Divergence}.\hskip 1em plus 0.5em minus
  0.4em\relax Springer Berlin Heidelberg, 2011, p. 720–722.

\bibitem{chawla2002smote}
N.~V. Chawla, K.~W. Bowyer, L.~O. Hall, and W.~P. Kegelmeyer, ``Smote:
  synthetic minority over-sampling technique,'' \emph{Journal of artificial
  intelligence research}, vol.~16, pp. 321--357, 2002.

\bibitem{Wan2022}
Y.~Wan, W.~Zhao, H.~Zhang, Y.~Sui, G.~Xu, and H.~Jin, ``What do they capture?''
  in \emph{Proceedings of the 44th International Conference on Software
  Engineering}.\hskip 1em plus 0.5em minus 0.4em\relax {ACM}, May 2022.

\bibitem{Xu2022}
F.~F. Xu, U.~Alon, G.~Neubig, and V.~J. Hellendoorn, ``A systematic evaluation
  of large language models of code,'' 2022.

\bibitem{Mehrabi2019}
N.~Mehrabi, F.~Morstatter, N.~Saxena, K.~Lerman, and A.~Galstyan, ``A survey on
  bias and fairness in machine learning,'' 2019.

\bibitem{lime}
M.~T. Ribeiro, S.~Singh, and C.~Guestrin, ``"why should {I} trust you?":
  Explaining the predictions of any classifier,'' in \emph{Proceedings of the
  22nd {ACM} {SIGKDD} International Conference on Knowledge Discovery and Data
  Mining, San Francisco, CA, USA, August 13-17, 2016}, 2016, pp. 1135--1144.

\bibitem{rebuffi2021}
S.-A. Rebuffi, S.~Gowal, D.~A. Calian, F.~Stimberg, O.~Wiles, and T.~Mann,
  ``Data augmentation can improve robustness,'' 2021.

\bibitem{rahman2023}
A.~M.~M. Rahman, W.~Yin, and G.~Wang, ``Data augmentation for text
  classification with {EASE},'' in \emph{Proceedings of the 6th International
  Conference on Natural Language and Speech Processing (ICNLSP 2023)}, M.~Abbas
  and A.~A. Freihat, Eds.\hskip 1em plus 0.5em minus 0.4em\relax Online:
  Association for Computational Linguistics, Dec. 2023, pp. 324--332.

\bibitem{na2023}
C.~Na, Y.~Choi, and J.-H. Lee, ``{DIP}: Dead code insertion based black-box
  attack for programming language model,'' in \emph{Proceedings of the 61st
  Annual Meeting of the Association for Computational Linguistics (Volume 1:
  Long Papers)}.\hskip 1em plus 0.5em minus 0.4em\relax Toronto, Canada:
  Association for Computational Linguistics, Jul. 2023, pp. 7777--7791.

\bibitem{Wang2021}
Y.~Wang, G.~Huang, S.~Song, X.~Pan, Y.~Xia, and C.~Wu, ``Regularizing deep
  networks with semantic data augmentation,'' \emph{IEEE Transactions on
  Pattern Analysis and Machine Intelligence}, p. 1–1, 2021.

\bibitem{cwe_cluster}
M.~Corporation, ``{C}{W}{E} - {C}{W}{E}-888: {S}oftware {F}ault {P}attern
  ({S}{F}{P}) {C}lusters (4.12) --- cwe.mitre.org,''
  \url{https://cwe.mitre.org/data/definitions/888.html}, [Accessed 28-07-2023].

\bibitem{rozière2024code}
B.~R. et~al., ``Code llama: Open foundation models for code,'' 2024.

\bibitem{jiang2024mixtral}
A.~Q.~J. et~al., ``Mixtral of experts,'' 2024.

\bibitem{dettmers2023qlora}
T.~Dettmers, A.~Pagnoni, A.~Holtzman, and L.~Zettlemoyer, ``Qlora: Efficient
  finetuning of quantized llms,'' 2023.

\bibitem{touvron2023llama}
H.~T. et~al., ``Llama 2: Open foundation and fine-tuned chat models,'' 2023.

\bibitem{Iraqi_2023}
\BIBentryALTinterwordspacing
M.~Iraqi, ``Comparing the performance of llms: A deep dive into roberta, llama
  2, and mistral for disaster tweets analysis with lora,'' Nov 2023. [Online].
  Available:
  \url{https://huggingface.co/blog/Lora-for-sequence-classification-with-Roberta-Llama-Mistral}
\BIBentrySTDinterwordspacing

\bibitem{grahn2021analysis}
D.~Grahn and J.~Zhang, ``An analysis of c/c++ datasets for machine
  learning-assisted software vulnerability detection,'' in \emph{Proceedings of
  the Conference on Applied Machine Learning for Information Security, 2021},
  2021.

\bibitem{russell2018automated}
R.~Russell, L.~Kim, L.~Hamilton, T.~Lazovich, J.~Harer, O.~Ozdemir,
  P.~Ellingwood, and M.~McConley, ``Automated vulnerability detection in source
  code using deep representation learning,'' in \emph{2018 17th IEEE
  international conference on machine learning and applications (ICMLA)}.\hskip
  1em plus 0.5em minus 0.4em\relax IEEE, 2018, pp. 757--762.

\bibitem{neuhaus2009beauty}
S.~Neuhaus and T.~Zimmermann, ``The beauty and the beast: Vulnerabilities in
  red hat's packages.'' in \emph{USENIX annual technical conference}, 2009, pp.
  527--538.

\bibitem{yan2017machine}
H.~Yan, Y.~Sui, S.~Chen, and J.~Xue, ``Machine-learning-guided typestate
  analysis for static use-after-free detection,'' in \emph{Proceedings of the
  33rd Annual Computer Security Applications Conference}, 2017, pp. 42--54.

\bibitem{hearst1998support}
M.~A. Hearst, S.~T. Dumais, E.~Osuna, J.~Platt, and B.~Scholkopf, ``Support
  vector machines,'' \emph{IEEE Intelligent Systems and their applications},
  vol.~13, no.~4, pp. 18--28, 1998.

\bibitem{quinlan1996learning}
J.~R. Quinlan, ``Learning decision tree classifiers,'' \emph{ACM Computing
  Surveys (CSUR)}, vol.~28, no.~1, pp. 71--72, 1996.

\bibitem{breiman2001random}
L.~Breiman, ``Random forests,'' \emph{Machine learning}, vol.~45, pp. 5--32,
  2001.

\bibitem{peterson2009k}
L.~E. Peterson, ``K-nearest neighbor,'' \emph{Scholarpedia}, vol.~4, no.~2, p.
  1883, 2009.

\bibitem{lomio2022just}
F.~Lomio, E.~Iannone, A.~De~Lucia, F.~Palomba, and V.~Lenarduzzi,
  ``Just-in-time software vulnerability detection: Are we there yet?''
  \emph{Journal of Systems and Software}, p. 111283, 2022.

\bibitem{freund1996experiments}
Y.~Freund, R.~E. Schapire \emph{et~al.}, ``Experiments with a new boosting
  algorithm,'' in \emph{icml}, vol.~96.\hskip 1em plus 0.5em minus 0.4em\relax
  Citeseer, 1996, pp. 148--156.

\bibitem{friedman2001greedy}
J.~H. Friedman, ``Greedy function approximation: a gradient boosting machine,''
  \emph{Annals of statistics}, pp. 1189--1232, 2001.

\bibitem{chen2016xgboost}
T.~Chen and C.~Guestrin, ``Xgboost: A scalable tree boosting system,'' in
  \emph{Proceedings of the 22nd acm sigkdd international conference on
  knowledge discovery and data mining}, 2016, pp. 785--794.

\bibitem{li2017software}
J.~Li, P.~He, J.~Zhu, and M.~R. Lyu, ``Software defect prediction via
  convolutional neural network,'' in \emph{2017 IEEE International Conference
  on Software Quality, Reliability and Security (QRS)}.\hskip 1em plus 0.5em
  minus 0.4em\relax IEEE, 2017, pp. 318--328.

\bibitem{cheng2019static}
X.~Cheng, H.~Wang, J.~Hua, M.~Zhang, G.~Xu, L.~Yi, and Y.~Sui, ``Static
  detection of control-flow-related vulnerabilities using graph embedding,'' in
  \emph{2019 24th International Conference on Engineering of Complex Computer
  Systems (ICECCS)}.\hskip 1em plus 0.5em minus 0.4em\relax IEEE, 2019, pp.
  41--50.

\bibitem{li2021sysevr}
Z.~Li, D.~Zou, S.~Xu, H.~Jin, Y.~Zhu, and Z.~Chen, ``Sysevr: A framework for
  using deep learning to detect software vulnerabilities,'' \emph{IEEE
  Transactions on Dependable and Secure Computing}, 2021.

\bibitem{lecun1995convolutional}
Y.~LeCun, Y.~Bengio \emph{et~al.}, ``Convolutional networks for images, speech,
  and time series.''

\bibitem{hochreiter1997long}
S.~Hochreiter and J.~Schmidhuber, ``Long short-term memory,'' \emph{Neural
  computation}, vol.~9, no.~8, pp. 1735--1780, 1997.

\bibitem{chung2014empirical}
J.~Chung, C.~Gulcehre, K.~Cho, and Y.~Bengio, ``Empirical evaluation of gated
  recurrent neural networks on sequence modeling,'' \emph{arXiv preprint
  arXiv:1412.3555}, 2014.

\bibitem{kipf2016semi}
T.~N. Kipf and M.~Welling, ``Semi-supervised classification with graph
  convolutional networks,'' \emph{arXiv preprint arXiv:1609.02907}, 2016.

\bibitem{scandariato2014predicting}
R.~Scandariato, J.~Walden, A.~Hovsepyan, and W.~Joosen, ``Predicting vulnerable
  software components via text mining,'' \emph{IEEE Transactions on Software
  Engineering}, vol.~40, no.~10, pp. 993--1006, 2014.

\bibitem{ding2024}
Y.~Ding, Y.~Fu, O.~Ibrahim, C.~Sitawarin, X.~Chen, B.~Alomair, D.~Wagner,
  B.~Ray, and Y.~Chen, ``Vulnerability detection with code language models: How
  far are we?'' 2024.

\bibitem{hagan1997neural}
M.~T. Hagan, H.~B. Demuth, and M.~Beale, \emph{Neural network design}.\hskip
  1em plus 0.5em minus 0.4em\relax PWS Publishing Co., 1997.

\end{thebibliography}
